\documentclass{jfm}

\usepackage{graphicx}
\usepackage{natbib}
\usepackage{multirow}
\usepackage{color}

\ifCUPmtlplainloaded \else
  \checkfont{eurm10}
  \iffontfound
    \IfFileExists{upmath.sty}
      {\typeout{^^JFound AMS Euler Roman fonts on the system,
                   using the 'upmath' package.^^J}%
       \usepackage{upmath}}
      {\typeout{^^JFound AMS Euler Roman fonts on the system, but you
                   dont seem to have the}%
       \typeout{'upmath' package installed. JFM.cls can take advantage
                 of these fonts,^^Jif you use 'upmath' package.^^J}%
      }
  \else
  \fi
\fi

\ifCUPmtlplainloaded \else
  \checkfont{msam10}
  \iffontfound
    \IfFileExists{amssymb.sty}
      {\typeout{^^JFound AMS Symbol fonts on the system, using the
                'amssymb' package.^^J}%
       \usepackage{amssymb}%

      }{}
  \fi
\fi

\ifCUPmtlplainloaded \else
  \IfFileExists{amsbsy.sty}
    {\typeout{^^JFound the 'amsbsy' package on the system, using it.^^J}%
     \usepackage{amsbsy}}
    {}
\fi

\newsavebox{\astrutbox}
\sbox{\astrutbox}{\rule[-5pt]{0pt} {20pt}}

\title[Accumulation of motile micro-organisms in turbulence]{Accumulation of motile elongated micro-organisms in turbulence}

\author[C. J. Zhan, G. Sardina, E. Lushi and L. Brandt]%
{C\ls A\ls I\ls J\ls U\ls A\ls N\ns  Z\ls H\ls A\ls N$^1$,
G\ls A\ls E\ls T\ls A\ls N\ls O\ns S\ls A\ls R\ls D\ls I\ls N\ls A$^{1,2}$
E\ls N\ls K\ls E\ls L\ls E\ls  I\ls D\ls A\ns L\ls U\ls S\ls H\ls I$^{3}$
\and\ls L\ls U\ls C\ls A\ns B\ls R\ls A\ls N\ls D\ls T$^1$%
\thanks{Email address for correspondence: luca@mech.kth.se}\ns
}

\affiliation{$^1$Linn\'{e} Flow Centre and SeRC (Swedish e-Science Research Centre),\\ KTH Mechanics, SE-100 44, Stockholm, Sweden \\[\affilskip]
$^2$Facolt\'a di Ingegneria, Architettura e Scienze Motorie, \\UKE Universit\'a Kore di Enna, 94100
Enna, Italy\\[\affilskip]
$^3$ School of Engineering, Brown University, \\182 Hope Street, Providence, Rhode Island 02912, USA\\ [\affilskip]
}

\pubyear{}
\volume{}
\pagerange{}

\begin{document}

\maketitle
\sloppy
\begin{abstract}
We study the effect of turbulence on marine life by performing numerical simulations of motile microorganisms, modelled as prolate spheroids, in isotropic homogeneous turbulence.
We show that the clustering and patchiness observed in {laminar flows, linear shear and vortex flows,} {are significantly reduced} in a three-dimensional turbulent flow mainly because of the complex topology; {elongated micro-orgamisms show some level of clustering in the case of swimmers without any preferential alignment whereas spherical swimmers remain uniformly distributed.}
Micro-organisms with one preferential swimming direction (e.g.\ gyrotaxis) still show significant clustering if spherical in shape, whereas prolate swimmers remain more uniformly distributed. Due to their large sensitivity to the local shear, these elongated swimmers react slower to the action of vorticity and gravity and therefore do not have time to accumulate in a turbulent flow. 
These results show how purely hydrodynamic effects can alter the ecology of microorganisms that can vary their shape and their preferential orientation.
\end{abstract}


\section{Introduction}

{The macroscopic phenomena of marine landscape are influenced by the interactions between the flow and the motility of
bacteria and phytoplankton. 
The effect of turbulence on marine life is therefore a key research question that also has relevance on the understanding of the consequences of climate changes. 
Microorganisms concentrate in the turbulent regions close to the surface and to the sea bed where the level of turbulence is also affected by external factors.}  The motion of an individual microorganism is determined by its swimming and by the advection of the fluid, where vorticity and rate of strain re-orient  it, and by the response to external stimuli and biases such as nutrient concentration, gravity and light. Depending on the external stimulus, the behavior is categorized for example as geotaxis \citep[]{Adams99}, phototaxis \citep[]{Martin83}, gyrotaxis \citep[]{Kessler85} and chemotaxis \citep[]{Adler74}, etc.
 
Among the different biases, we consider gyrotaxis. This bias results from the
combination of a viscous torque on the cell body, caused by the flow shear, and a
gravitational torque, arising from an asymmetric distribution of mass within the organism (bottom-heaviness).
This induces an accumulation of cells heavier then water at the free surface and the occurrence of a bioconvective instability that develops from an initially uniform suspension without any background flow \citep{Pedley92,Pedley93}. 

Recent studies consider the interactions between a complex flow and swimming microorganisms.
Cellular laminar flows are used by  \cite{Torney07} to study the aggregation of self-propelled particles without taxis: these authors 
show that the particles concentrate around {chaotic trajectories}. 
Clustering is more pronounced for prolate swimmers and higher swimming speeds when particles escape from regular
elliptic regions. In this particular case, spherical particles cannot enter these regions {if initially outside of them}.
The accumulation of swimming microorganisms in chaotic
regions of a fluid flow can be advantageous for fast
dispersion. \cite*{Khurana11} consider a two-dimensional chaotic vortical flow and show that swimming does not necessarily lead to enhanced particle transport. For small but finite values of the
swimming speed, particles can be trapped for very long time near the boundaries between chaotic and regular flow region.
In a later study, \cite{Khurana12} add stochastic terms to the swimmer equation of motion and notice that suppression of transport (trapping) will increase dramatically with the addition of these random motions and rotational stochasticity and, more significantly, with elongated particles. 
At higher swimming speeds, elongated
swimmers tend to be attracted to the stable manifolds of hyperbolic fixed points, leading
to increased transport relative to swimming spheres.

The coupling of gyrotactic particles and a layer with higher shear is studied by 
\cite*{Durham09} who demonstrated that motility and shear are responsible for the formations of intense thin layers of phytoplankton by gyrotactic trapping. 
These layers appear in regions characterized by a vertical gradient of the horizontal velocity that exceeds a critical threshold value: cells cluster in thin layers  and  tumble end over end.
The coupling between the flow and the swimmers' motility may lead to the formation of macroscopic flow features in the more general case of more complex flows. \citet*{Durham11} investigated how the gyrotactic motile microorganisms aggregate in a steady TGV flow (Taylor-Green Vortex) and suggested that the patchiness regimes can be characterized by two non-dimensional parameters: the swimming speed relative to a characteristic fluid speed and the magnitude of the gyrotactic torque.

In this paper, we will consider the behavior of gyrotactic and non-gyrotactic swimmers in a three-dimensional turbulent flow, as turbulence characterizes the life of microbes in water supply systems, ocean and bioreactors. As in previous studies, we approximate swimmers as prolate spheroids. The statistics of  non-swimming ellipsoidal particles in the turbulent flow can be found in \cite{Parsa12} and references therein.  These studies show that the rotation rate is influenced by alignment, particles orientations become correlated with the velocity gradient tensor, and the alignment depends strongly on the particle shape. 
\cite{Lewis03} proposes a Fokker-Planck model for the orientation of motile spheroids in homogeneous isotropic turbulence and shows that fluctuations in orientation manifest themselves as an increase of the effective rotary diffusivity.
\cite{thorn} considered the long-time trajectories of deterministic and stochastic swimmers in shear flows. These authors simulated stochastic gyrotactic micro-organisms in synthetic turbulence and demonstrate quantitatively the transition between swimming-dominated drift and turbulence-dominated diffusion as a function of the kinetic energy dissipation rate.
\cite*{Lillo12} describe the spatial distribution of gyrotactic spherical microorganisms transported by three-dimensional turbulence flows generated by DNS (Direct Numerical Simulations). They show that coupling gyrotactic motion to turbulent flow produces small-scale patchiness (smaller than the Kolmogorov scale) in the swimmer distribution. 
\citet[]{Durham12,Durham13} also examined whether gyrotaxis can generate cell patchiness in turbulent flow using experiments and 
numerical simulations.
They found that accumulation in downwelling regions is the dominant means of aggregation also in turbulent flow and shows that patchiness is not significantly affected by the Taylor Reynolds number $Re_{\lambda}$.
\cite{Croze12} studied dispersion of gyrotactic algae in turbulent channel flow and quantify the increased diffusivity induced by the turbulent fluctuations.

In this paper, we document how clustering of prolate swimmers without taxis, observed in simple flows, is destroyed by turbulence. We then consider gyrotactic microorganisms of different shapes and confirm that bottom-heavy microorganisms accumulate in downwelling flows.
This explains how settling larvae changing the offset of the centers of buoyancy and of gravity can preferentially accumulate in updrafts, favorable for dispersal, or downdrafts, favorable for settlement, thus exploiting the hydrodynamics of the vorticity near the sea bed \citep{grunbaum03}.
We finally show  that clustering is most evident for spherical shapes. This suggests that micro-organisms like the dinoflagellate \textit{Ceratocorys horrida}, able to reversibly change its morphology in response to variations of the ambient flow \citep*{Zirbel02}, can exploit hydrodynamics effects to increase or decrease encounter rates in an active way.

\section{Problem Formulation}\label{sec:problem-formulation}
\subsection{Governing equations}
The flow velocities are solution of the Navier-Stokes equation
\begin{equation}
\frac{\partial \mathbf{u}}{\partial t}+\left(\mathbf{u}\cdot \nabla\right)\mathbf{u}=-\nabla p+\frac{1}{Re} \nabla ^2 \mathbf{u}+\mathbf{f},
\label{eq1}
\end{equation}
\begin{equation}
\nabla \cdot \mathbf{u}=0.
\label{eq2}
\end{equation}
Here, $\mathbf{u}$ is the fluid velocity, $p$ is the pressure and $\mathbf{f}$ is the external large-scale forcing needed to keep homogeneous isotropic turbulence in three-periodic domain. $Re$ is the Reynolds number.  

The swimmers in our simulation are prolate spheroids advected by the local velocity while moving with constant speed $u_{s}$
\begin{equation}
\frac{d\mathbf{x}}{dt}= \mathbf{u} +u_{s} \mathbf{p},
\label{eq3}
\end{equation}
where the versor $\mathbf{p}$ defines the orientation of the swimmers. Assuming inertialess motion, the angular velocity of the organisms is determined by the balance of viscous and gyrotactic torques \citep{Pedley92}
\begin{equation}
\frac{d\mathbf{p}}{dt}=\frac{1}{2B} \left[ \mathbf{k}-\left( \mathbf{k}\cdot \mathbf{p}\right) \mathbf{p}\right]+\frac{1}{2}\mathbf{\omega} \times \mathbf{p}+
\alpha\left[I-\mathbf{pp}\right] \cdot \mathbf{E} \cdot \mathbf{p}.
\label{eq4}
\end{equation}
In the above equation, $\mathbf{E}$ is symmetric part of the deformation tensor and $\mathbf{\omega} $ is the vorticity vector. $\alpha =\left(\mathcal{AR}^2-1\right)/\left(\mathcal{AR}^2+1\right)$ defines the eccentricity of the spheroids and $\mathcal{AR}$ the corresponding aspect ratio (the ratio of the major to the minor axis). $B$ is the characteristic time a perturbed cell takes to return to the vertical orientation, $\mathbf{k}$, the preferred swimming direction (e.g. gyrotaxis).  When there is no preferred direction, $B\to \infty$ and the first term on the right hand side vanishes. 

\subsection{Numerical methodology}
The numerical data set has been obtained from a Direct Numerical Simulation
(DNS)  using a classical pseudo-spectral method coupled with a Lagrangian solver
for the swimming micro-organisms.
For the fluid phase, the Navier-Stokes equations have been integrated in a 
three-periodic domain of length $L_D=2\pi$ using a Fourier spectral method with the nonlinear terms
de-aliased by the $3/2$ rule. The solution is advanced in time using a third-order
low-storage Runge-Kutta method; specifically, the nonlinear terms are computed
using an Adam-Bashforth like approximation while the diffusive terms are
analytically integrated \citep{rogallo}.
A random forcing is applied isotropically to the first shell of  wave vectors,
with fixed amplitude $\hat{f_0}$, constant in time and uniformly distributed in
phase and directions \citep{vinmen}.
We use a resolution of $192$ Fourier modes in each of the three
directions that correspond to a grid size in physical space of $288^3$ collocation points due to the de-aliasing.
The ratio between the highest resolved wave number
$k_{max}$ and the Kolmogorov scale $k_\eta$ is $k_{max}/k_\eta=1.73$, which is within the usual
accepted range to ensure a stable code and a good resolution \citep{pope}.
The simulation has a Taylor Reynolds number $Re_\lambda=150$.
To compute statistics, we analyze about $160$ configurations, stored 
{with an interval corresponding to half of the turbulence integral time 
scale} (for each case considered). 

For the dispersed phase, we use the point particle approximation as typical micro-organisms are smaller than the flow Kolmogorov length.
The same Runge-Kutta temporal integration used
for the carrier phase is adopted for the swimmers. The integration of Eq. (\ref{eq4}) for the orientation 
is performed using the quaternion formulation. A second-order interpolation scheme
is used to compute the flow velocity, vorticity and velocity gradients at particle position with a high enough resolution to capture
the flow velocity gradients. The swimmers are initially uniformly distributed in
the domain and reach a statistically steady state after a short transient.
Several simulations have been performed by keeping fixed the properties of the turbulence and changing the parameters of the swimming organisms, i.e.\  aspect ratio $\mathcal{AR}$ , swimming velocity and the reorientation time $B$. A list of the different simulations  is reported in Table \ref{tab:gyro}. In our simulation, the number of particles is 200,000. In the following, $V_{s}^*=u_{s}/u_{\eta}$ is the swimming velocity made non-dimensional with the Kolmogorov velocity scale and $\omega_{rms}$ the root mean square of the vorticity fluctuations.

\begin{table}
  \begin{center}
  \begin{tabular}{|c|c|p{5.5cm}|}
   $\mathcal{AR}$ & $V_{s}^*$ & $B\omega_{rms}$ \\ \hline \hline
    1 & 1 & 0.5, 1, 3, 5, $\infty$ \\[3pt] \hline
    3 & 1 & 1, $\infty$ \\[3pt]  \hline
    \multirow{5}{*}{9} & 0 & $\infty$ \\ 
     & 0.5 & $\infty$ \\
     & 1 & 0.2, 0.5, 1, 1.5, 2, 2.5, 3, 5, 10, 30, $\infty$ \\
     & 2 & $\infty$ \\
     & 5 & $\infty$ \\[3pt] \hline
    $\infty$ & 1 & 1, $\infty$ \\ \hline
  \end{tabular}
  \caption{Parameters defining the swimmers in the different simulations presented here. These evolve in a homogeneous isotropic turbulent flow with 
  $Re_\lambda=150$. $B\omega_{rms} = \infty$ indicates no gyrotaxis.}
  \label{tab:gyro}
  \end{center}
\end{table}

\section{Results}\label{sec:results}

\subsection{Non-gyrotactic swimmers}

\begin{figure}
\begin{center}
      \includegraphics[width=0.48\textwidth]{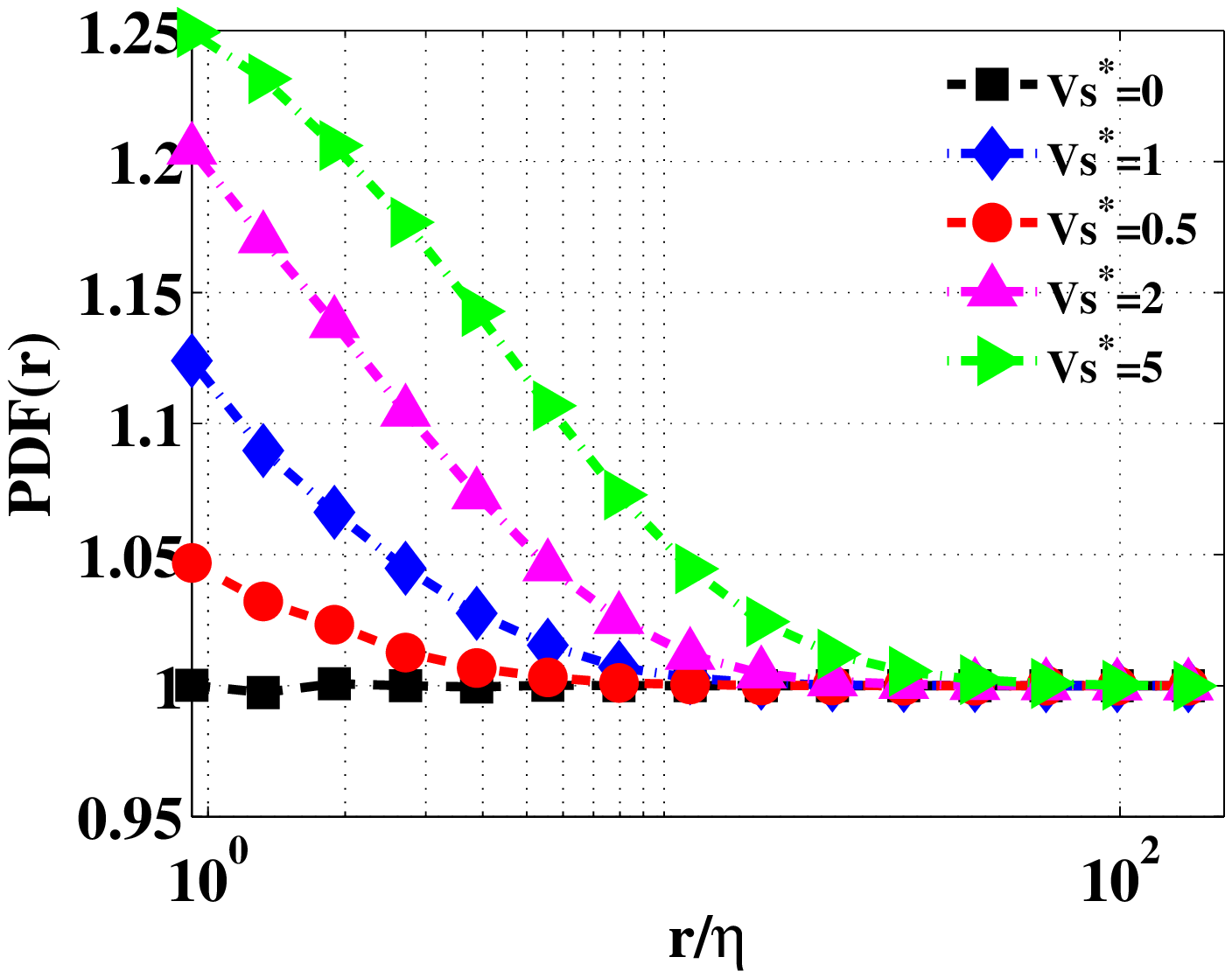}
      \put(-194,124){ $\left(a\right) $}
      \includegraphics[width=0.48\textwidth]{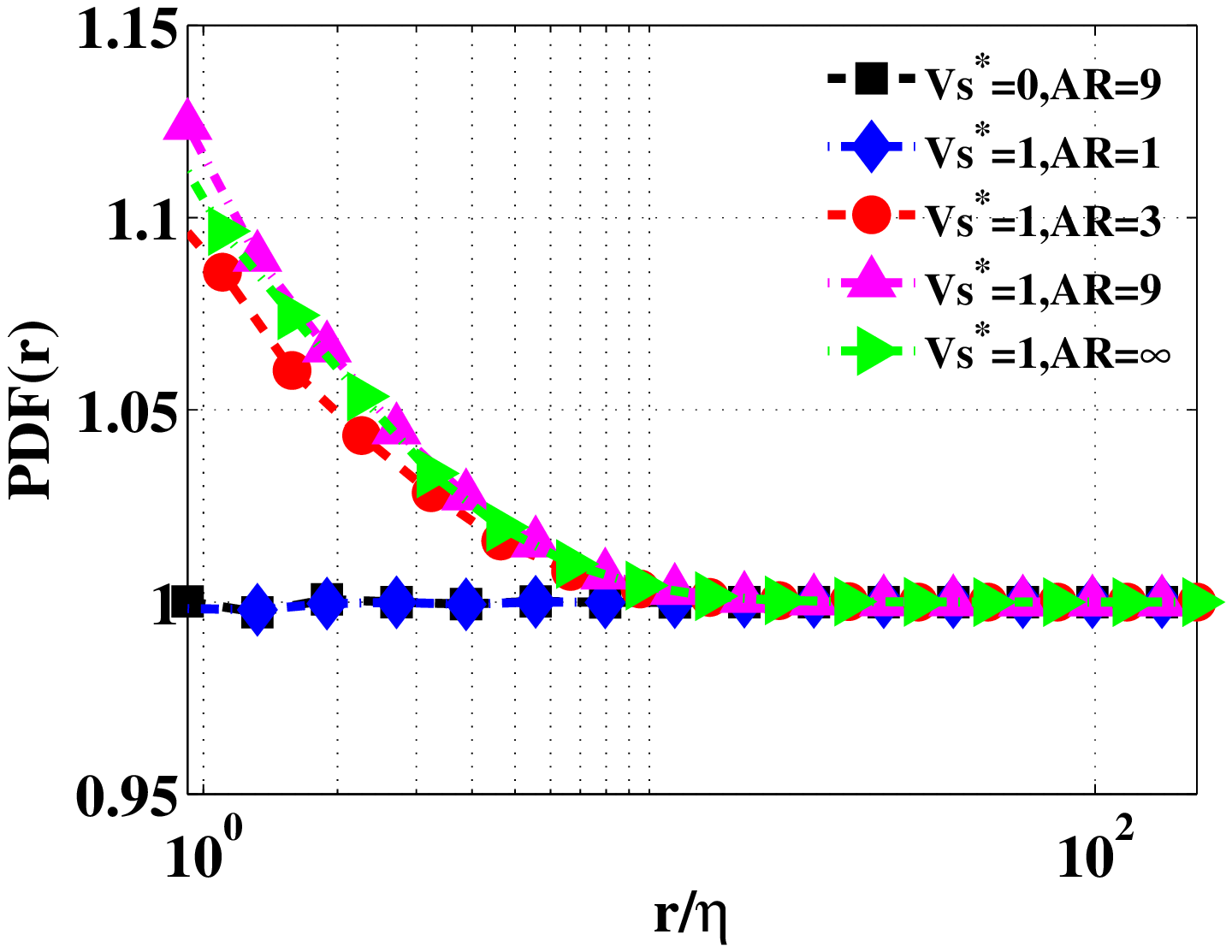}
       \put(-194,124){ $\left(b\right) $}
    \caption{Radial distribution function of particle pair for $\left(a\right) $ swimmers of same aspect ratio $\mathcal{AR}=9$ and different swimming speeds. $\left(b\right) $
    same swimming speed $V_{s}^*$=1 and different aspect ratios.}\label{fig:rdf}
\end{center}
\end{figure}

We first study the behavior of micro-organisms who do not possess a preferential swimming direction. These have been shown to accumulate and get trapped in simple vortical flows \citep[]{Torney07, Khurana11}. To quantify patchiness (clustering) in fully three-dimensional isotropic turbulent flows,  we use the radial pair distribution function (RDF), sometimes also called correlation function. This is defined as
\begin{equation}
g\left(r\right)=\frac{1}{4\pi r^2} \frac{dN_{r}}{dr} \frac{1}{n_{0}},
\label{eq5}
\end{equation}
where, $n_{0}=0.5N_{p}\left(N_{p}-1\right)/V_{0}$ is the density of pairs in the whole volume $V_{0}$. $N_{p}$ is the total number of particles in the domain and $N_{r}$ is the number of pairs at distance $r$. The RDF measures the probability to find a particle pair at a given radial distance normalized by the values of a uniform distribution. An indicator of patchiness is also the scaling exponent of RDF at small separations. 

Results for swimmers of different shape and different swimming speed are presented in figure~\ref{fig:rdf}.
Clustering, the value of the RDF at small separations,  is relatively weak and flow visualizations indeed show an almost uniform distribution.
The slope of the RDF at $r \to 0 $ is larger for higher swimming speeds and elongated particles: swimmers with $V_{s}^*=5$, $AR=9$ exhibit maximum accumulation. 
As expected, the populations characterized by zero swimming speed or spherical shape do not present any clustering as their velocity field is divergence free. In the first case they are advected as passive tracers, whereas in the second case the swimming orientations are uniformly distributed.
Large swimming velocities are able to counteract the dispersive effect of turbulence \citep{Croze12} and more elongated particles are associated to a potentially compressible velocity field,  being sensitive to the background shear.
  In summary, the accumulation is very weak and from our observations we can conclude that {in three-dimensional turbulence 
 patchiness exceeds that of a Poisson distribution only for elongated cells, not for spherical ones. This fact could be quite important for the swimmers' ecology and deserves further analysis.}

By tracking the swimmers in a frozen (time-independent) velocity field we also registered a significant decrease of the cell accumulations,  when compared to laminar flows. In particular, the value of the RDF at small $r$ increases only from 1.1 to 1.2 when removing the flow unsteadiness for the simulation with $\mathcal{AR}=9$. We therefore conclude that the complex three-dimensional flow topology is the main responsible of the lack of any significant clustering reported above.

\begin{figure}
\begin{center}
                \includegraphics[width=0.48\textwidth]{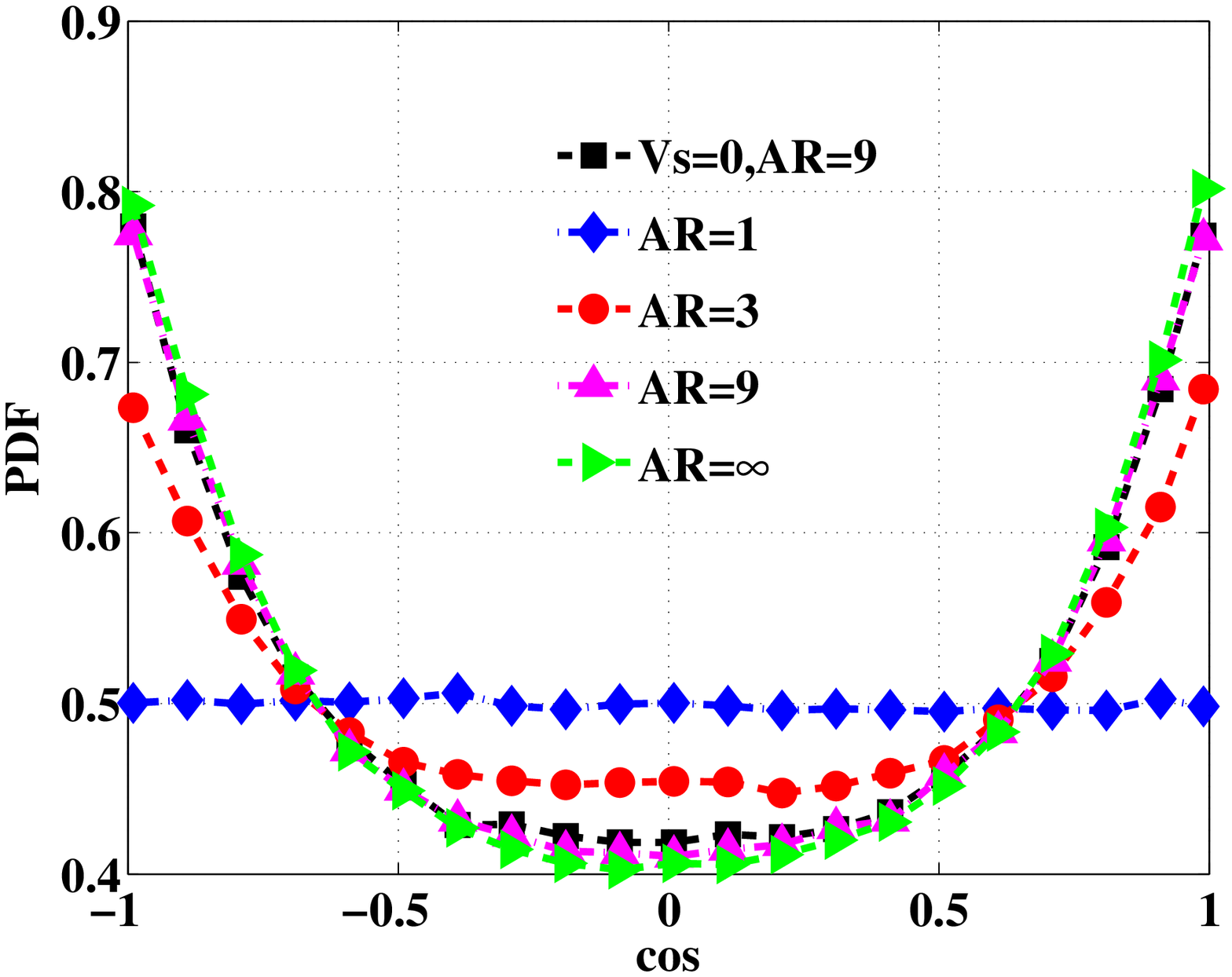}  \put(-192,139){ $\left(a\right) $}   \put(-79,2){$\theta_1$}   
               \hspace{10pt}  \includegraphics[width=0.48\textwidth]{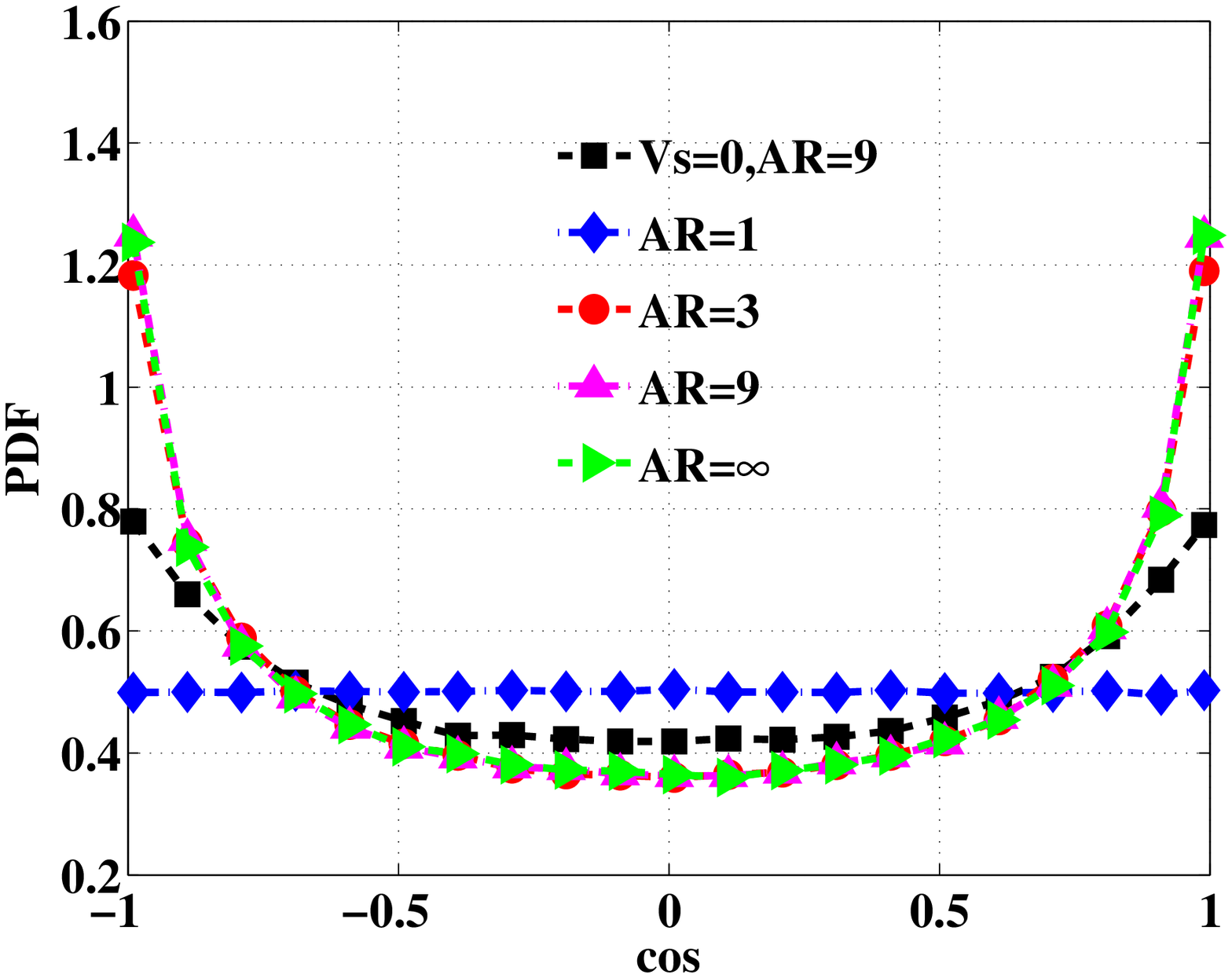}  \put(-192,139){ $\left(b\right) $}  \put(-79,2){$\theta_2$} 

                \includegraphics[width=0.48\textwidth]{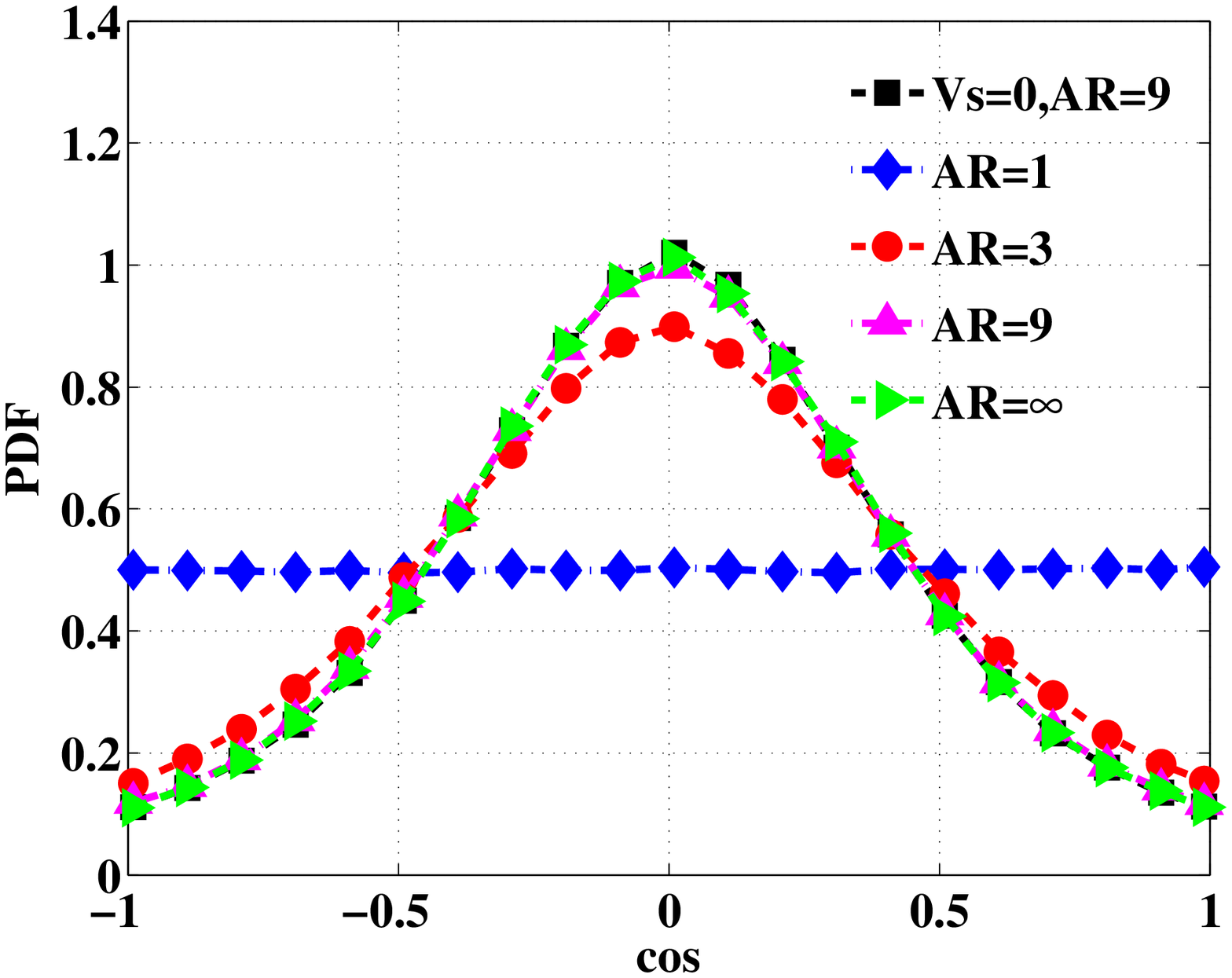}  \put(-192,139){ $\left(c\right) $} \put(-79,2){$\theta_3$} 
               \hspace{10pt}  \includegraphics[width=0.48\textwidth]{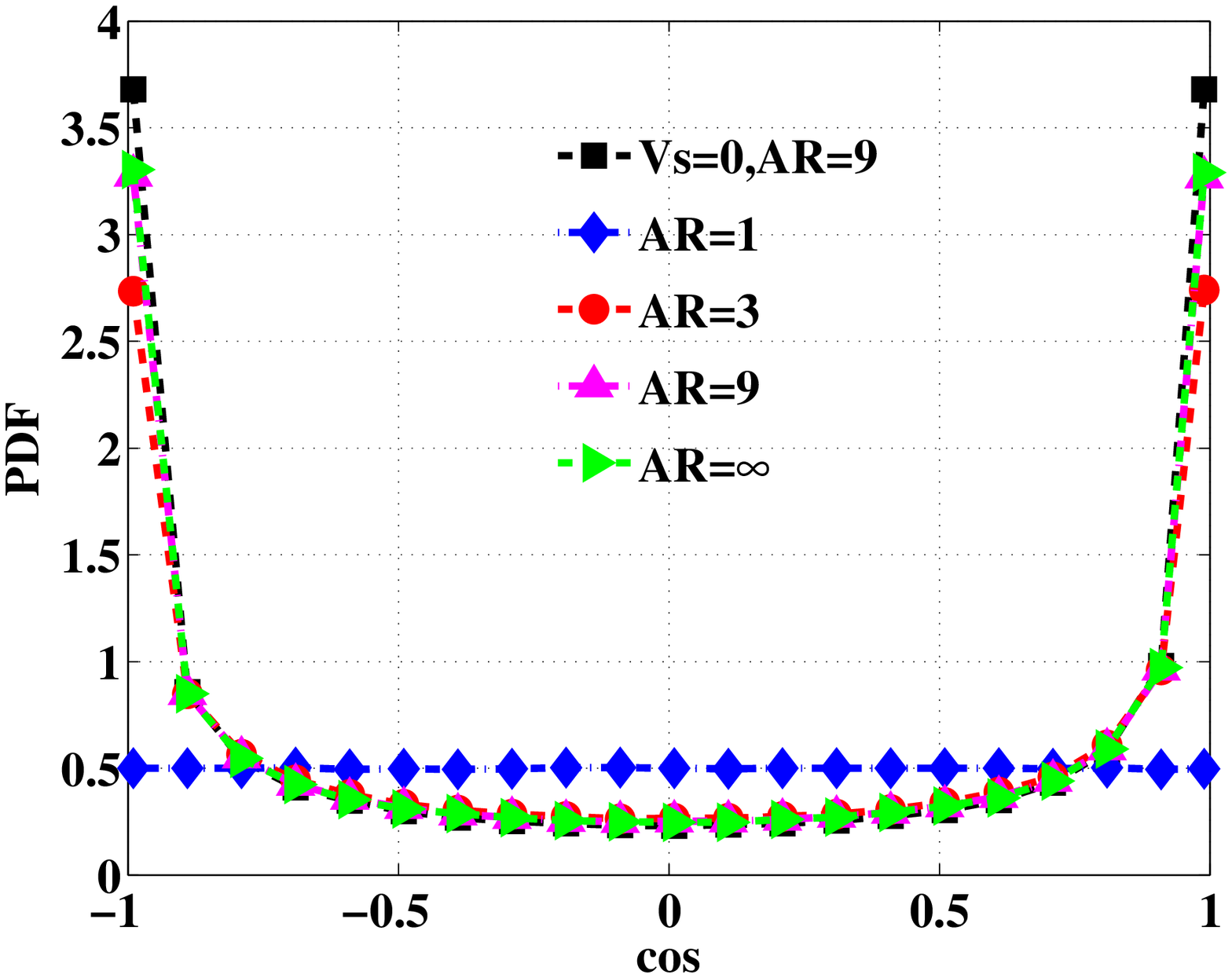}  \put(-192,139){ $\left(d\right) $} \put(-79,2){$\psi$} 
        \caption{PDF of orientation of swimmers with $V_{s}^*$=1 with respect to $\left(a\right) $, $\left(b\right) $ and $\left(c\right)$ the three eigenvectors of the deformation tensor; and $\left(d\right) $ the local vorticity vector}\label{fig:PDF_nogyro}
\end{center}
\end{figure}

In figure~\ref{fig:PDF_nogyro}, we aim to understand the relation between particle orientation and the underlying flow field when varying the micro-organism shape. 
In figure~\ref{fig:PDF_nogyro}(a),(b) and (c), we show the orientation with respect to the three eigendirections of the strain tensor, $\lambda_1> \lambda_2> \lambda_3$. Here, the angles between these eigendirections and the cell orientation are denoted as $\theta_1$, $\theta_2$ and $\theta_3$.
 $\psi$ represents the angle between the orientation and vorticity vector (figure~\ref{fig:PDF_nogyro}d).
 The spherical swimmers do not show a preferential orientation with strain or vorticity as expected by the lack of any accumulation.
 The pdfs show peaks when prolate swimmers are parallel to the eigendirections associated to $\lambda_1$ and $\lambda_2$. An increasing probability of parallel alignment with the  first eigendirection of the strain is seen at larger aspect ratio.  The swimming direction is more likely to be normal to the third eigendirection of the strain as shown in figure~\ref{fig:PDF_nogyro}(c). 
Interestingly, figure~\ref{fig:PDF_nogyro}(d) shows that the strongest tendency is to align with the local vorticity vector, which is the dominant effect.
This finding resembles previous observations in oscillatory flows of dense fiber suspensions \citep{Franceschini11}.  Non-motile {fibers}  are more likely to align with the vorticity vector than motile swimmers, whereas it is less likely to find them parallel to the second eigenvector of the strain tensor. 
The distributions of orientation with respect to the vorticity vector appears to weakly vary with the aspect ratio for $\mathcal{AR}>1$.

\begin{figure}
\begin{center}
               \includegraphics[width=0.4\textwidth]{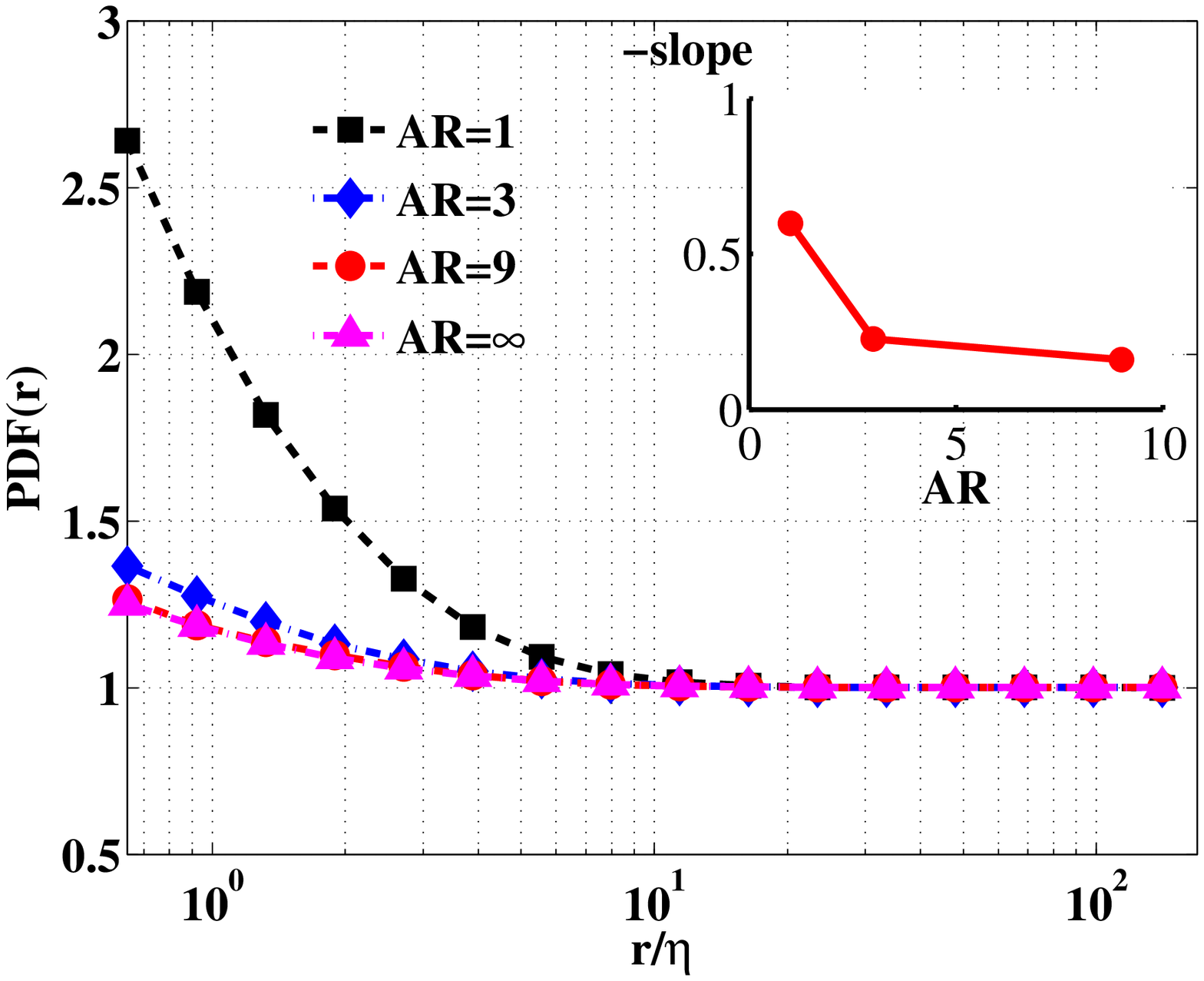} \put(-162,116){ $\left(a\right) $}
               \hspace{20pt} \includegraphics[width=0.4\textwidth]{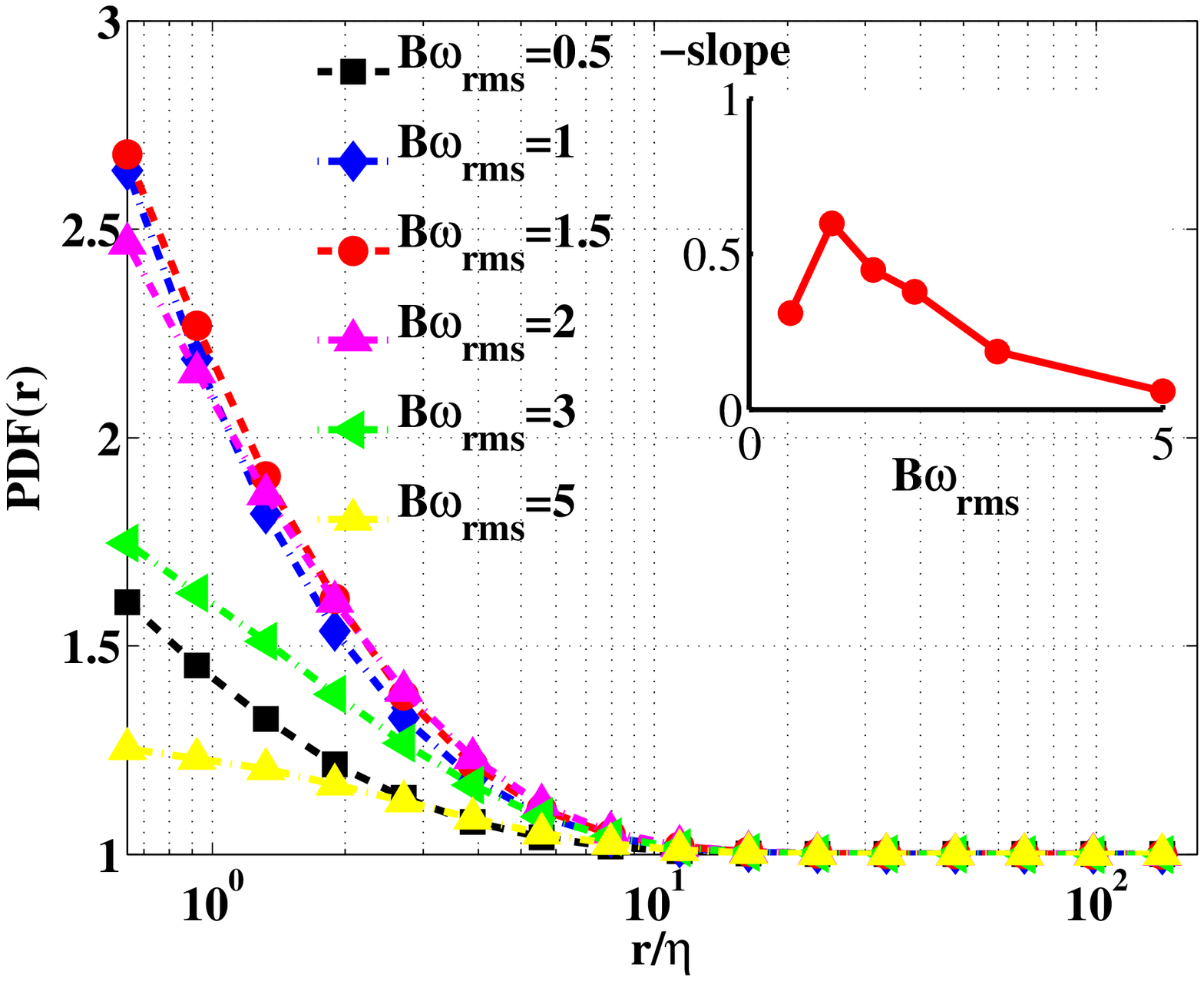} \put(-162,116){ $\left(b\right) $}
               
                \includegraphics[width=0.55\textwidth]{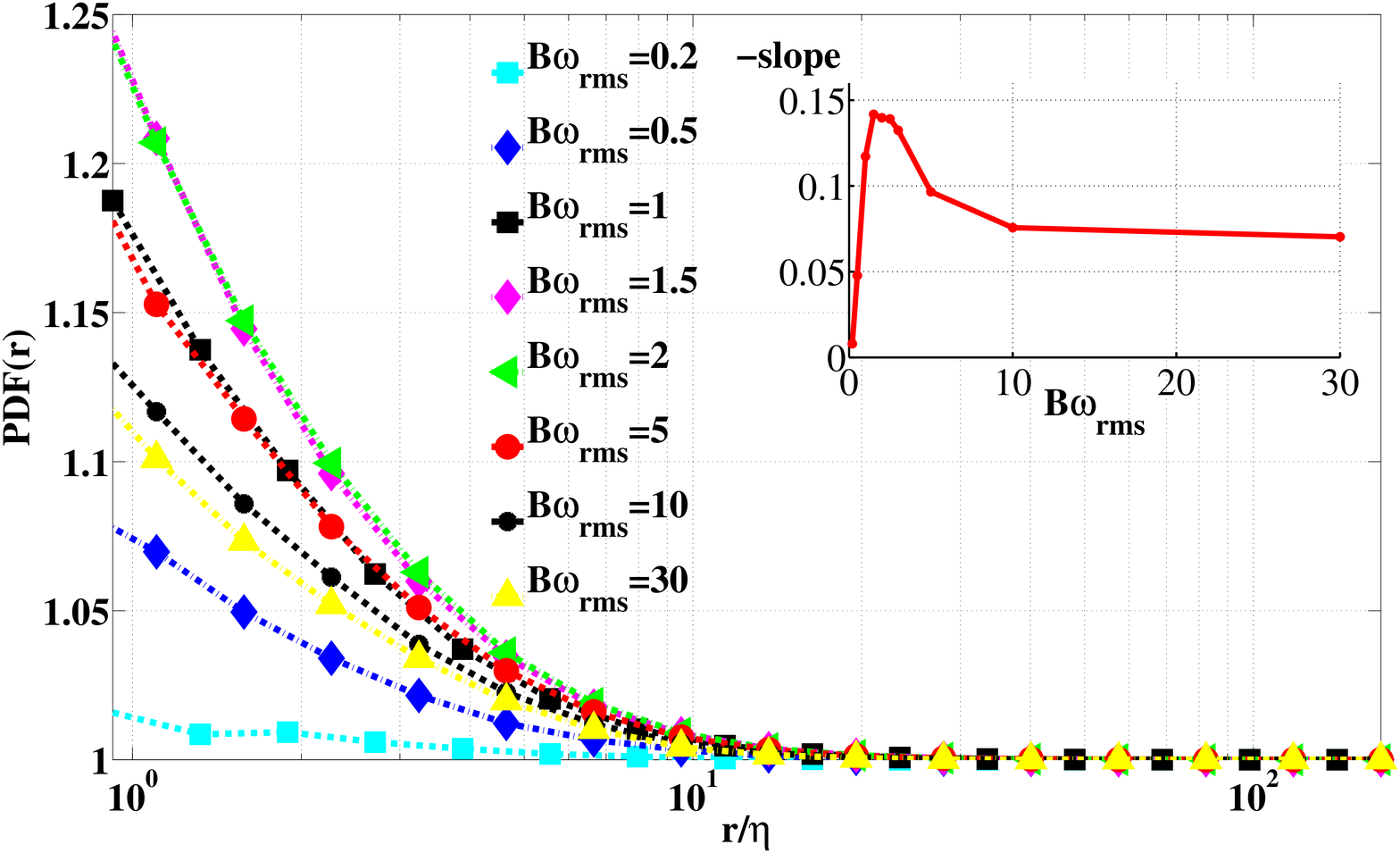} \put(-214,100){ $\left(c\right) $}
                \includegraphics[width=0.37\textwidth]{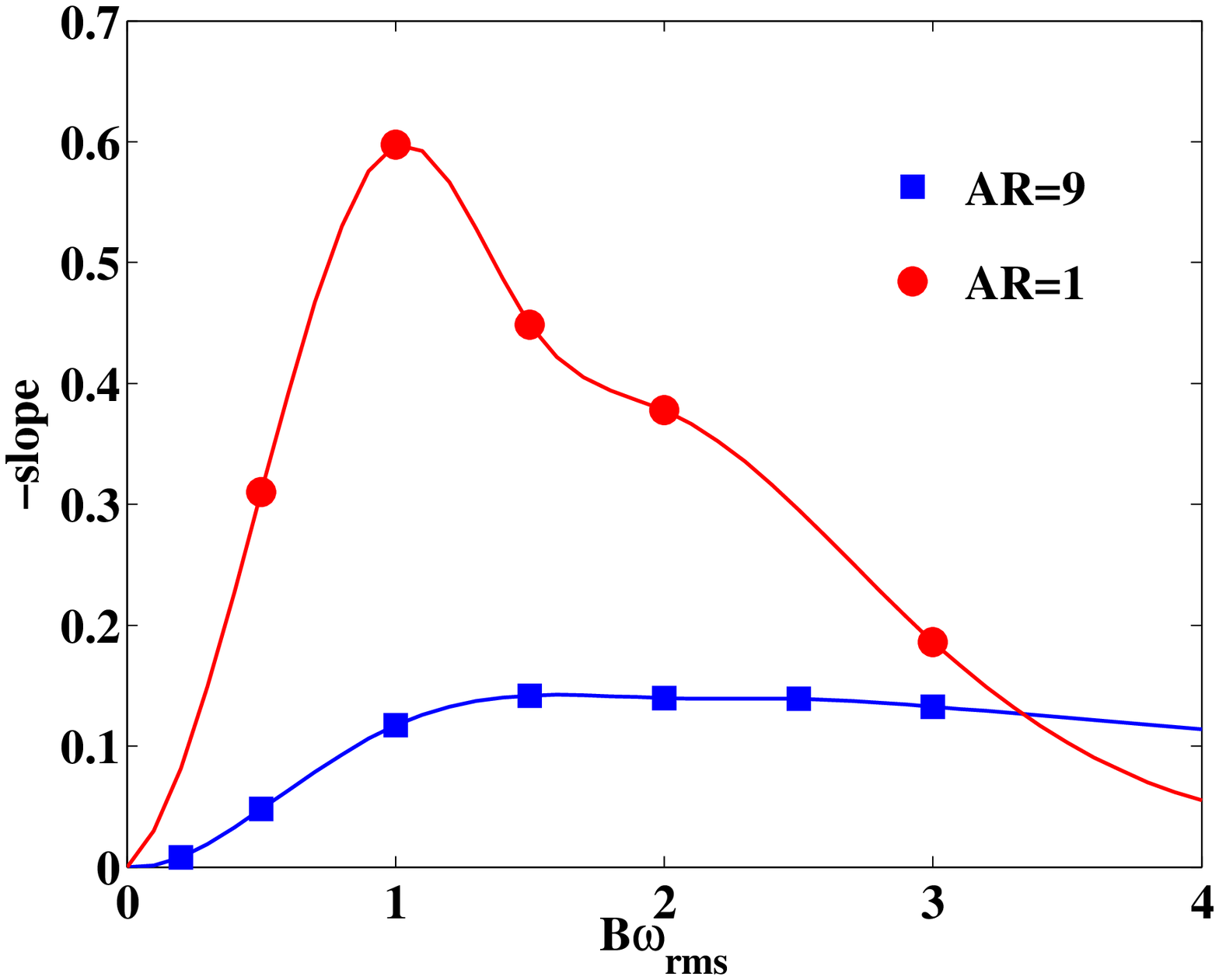} \put(-151,100){ $\left(d\right) $}

        \caption{Radial distribution function (RDF) for gyrotactic swimmers. $\left(a\right) $ $B\omega_{rms}=1$ and different aspect ratios as indicated. $\left(b\right) $ Aspect ratio $\mathcal{AR}=1$ and $\left(c\right) $ Aspect ratio $\mathcal{AR}=9$ and gyrotactic torque $B\omega_{rms}$ as indicated in the legend for the same swimming speed $V_{s}^*$=1. The inset displays the scaling exponent at small separations. {$\left(d\right) $ Close-up of the scaling exponent at small separations for the values of $B\omega_{rms}$ with maximum clustering.}
         {The slope of the RDF is estimated by fitting with an ordinary least square method a power law in the range r/$\eta$=[0.2:2].}}\label{fig:rdfgyro}
\end{center}
\end{figure}

\subsection{Gyrotactic swimmers}

Here we study bottom-heavy swimmers that tend to align with gravity and swim upwards and focus on the effect of the micro-orgamism shape. These type of swimmers are known to accumulate in regions of negative vertical velocity, in downwelling flows, also in the turbulent regime when spherical \citep{Kessler85,Durham12,Lillo12,Croze12}.
The main findings of this work are reported in figure~\ref{fig:rdfgyro}:
For gyrotactic swimmers of different shape clustering decreases when the aspect ratio of swimmers is increasing. In other words, spherical particles having a preferential swimming direction exhibit the most significant aggregation. 
Results in figure~\ref{fig:rdfgyro}(a) are obtained with fixed value of $B\omega_{rms}$, a non-dimensional parameter measuring the ratio of the alignment timescale to the rotation timescale induced by vorticity  \citep{Durham12}. 

\begin{figure}
\begin{center}
                \includegraphics[width=0.48\textwidth]{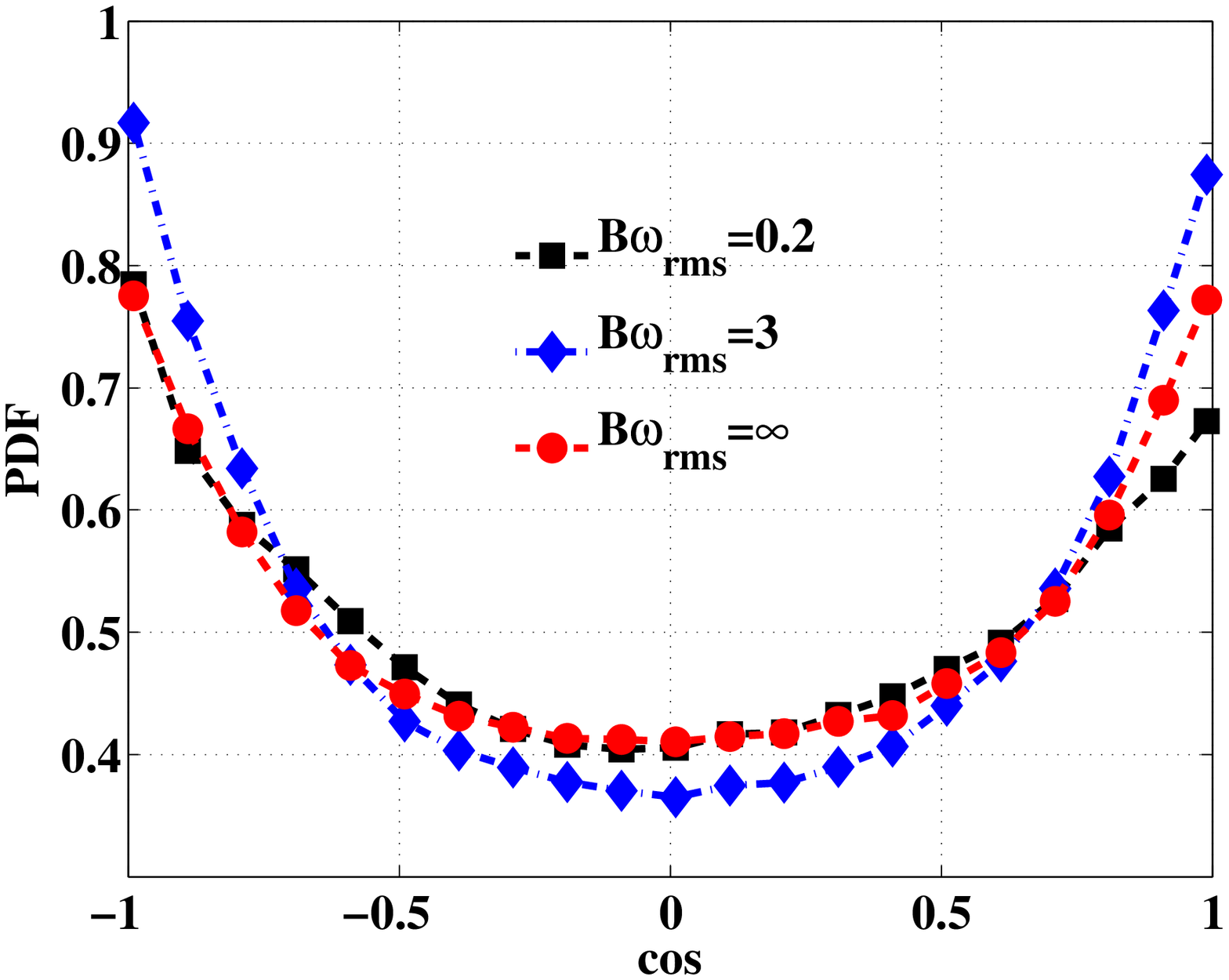} \put(-192,138){ $\left(a\right) $} \put(-79,2){$\theta_1$}
              \hspace{10pt}   \includegraphics[width=0.48\textwidth]{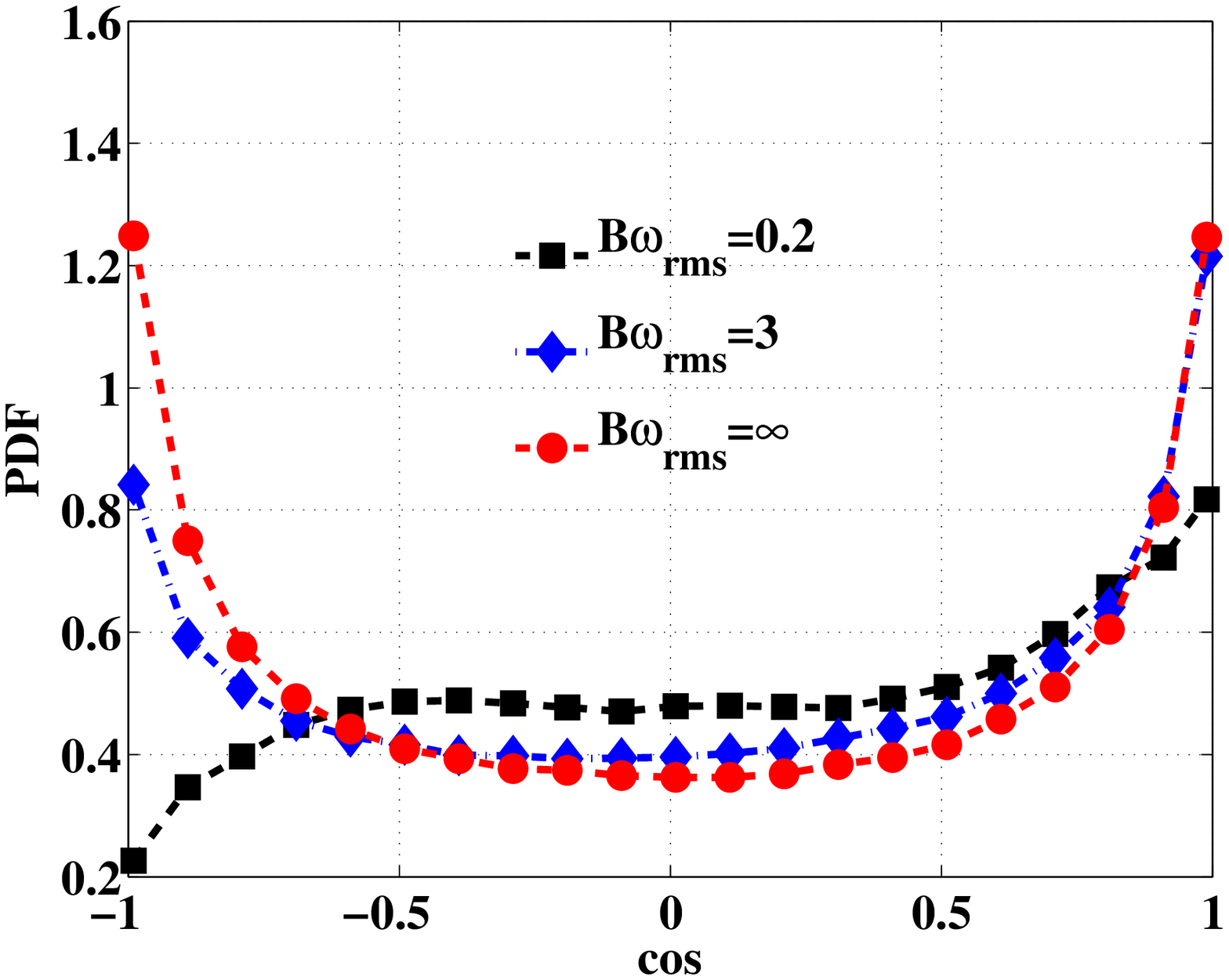}  \put(-194,138){ $\left(b\right) $} \put(-79,2){$\theta_2$}

                \includegraphics[width=0.48\textwidth]{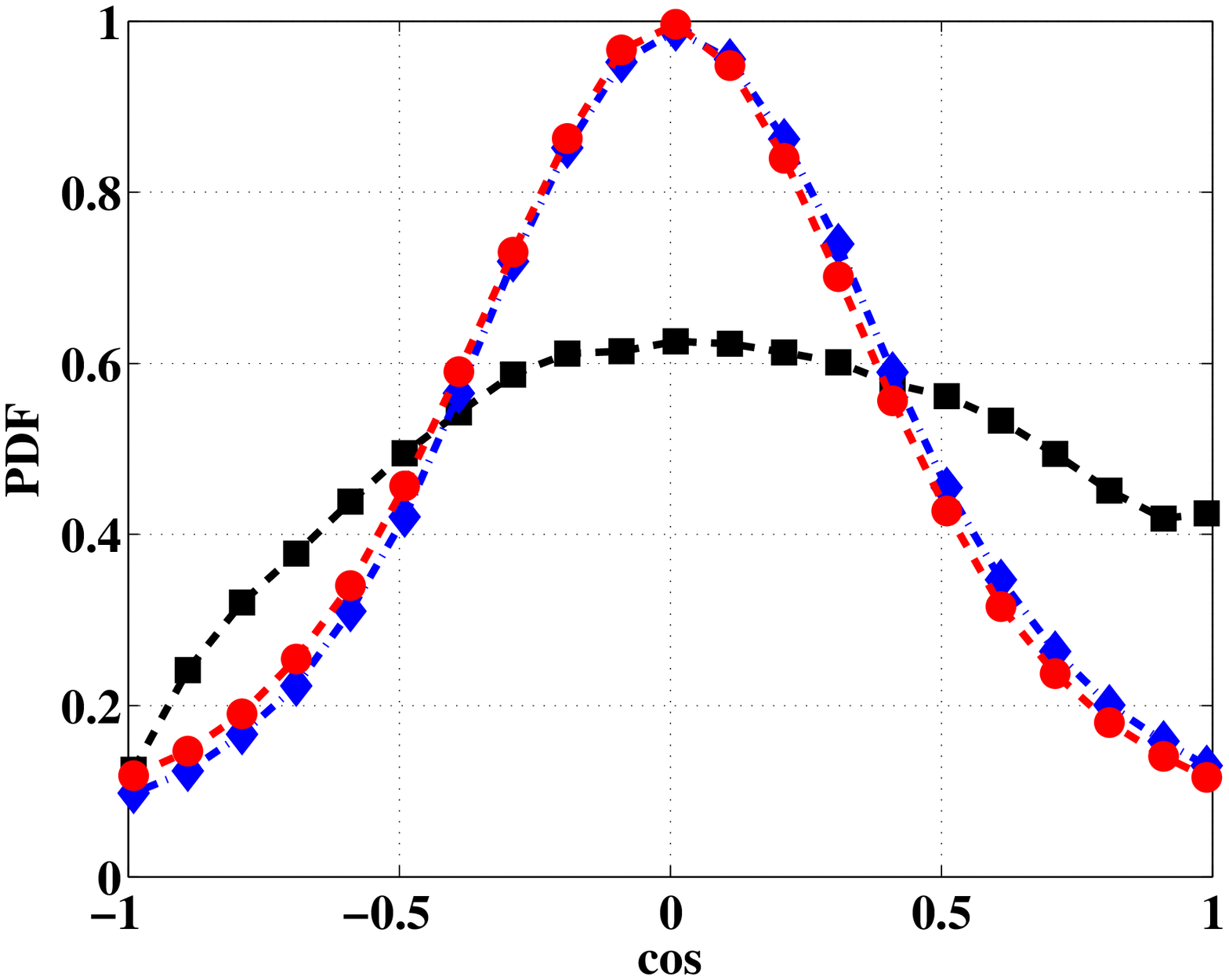}   \put(-192,138){ $\left(c\right) $} \put(-79,2){$\theta_3$}
              \hspace{10pt}   \includegraphics[width=0.48\textwidth]{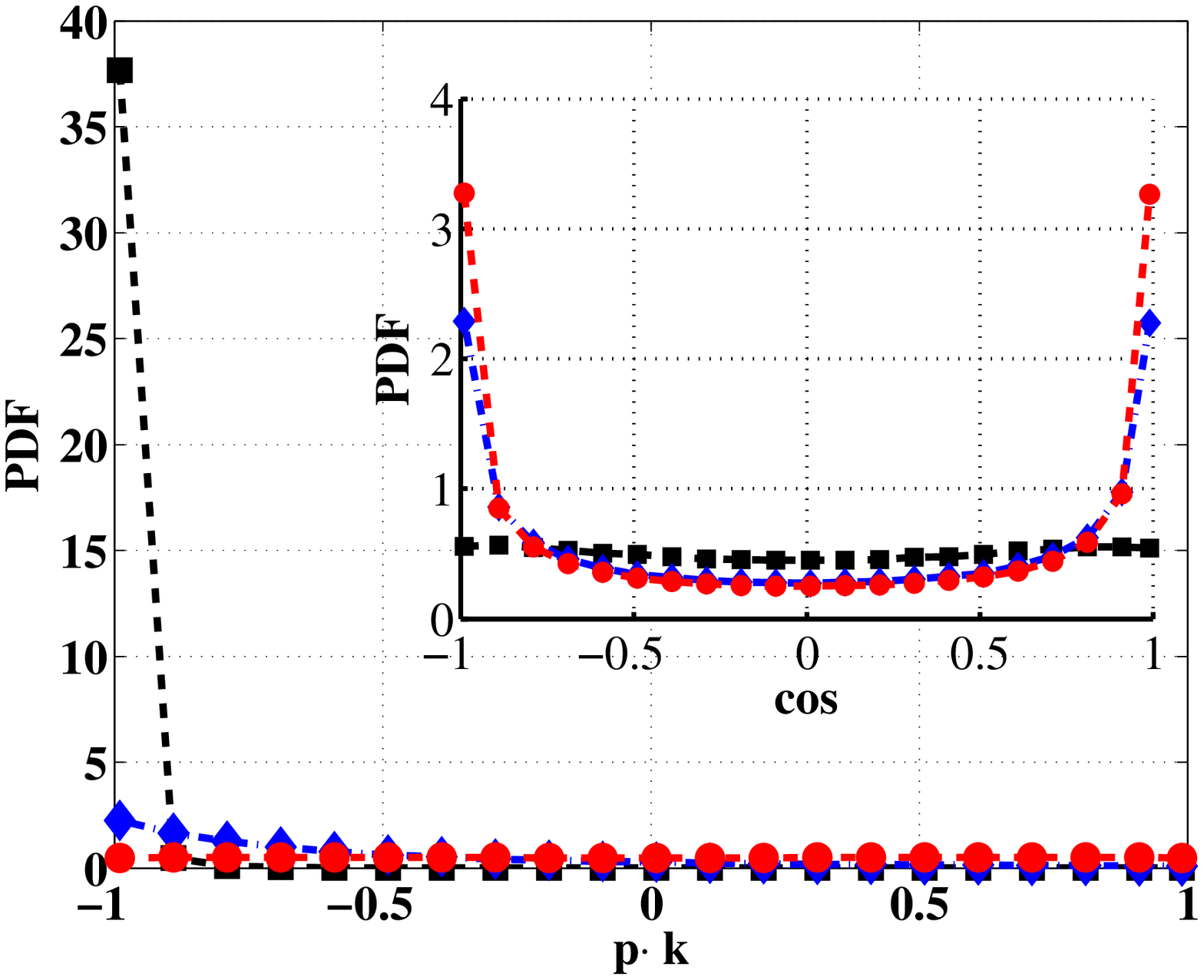}   \put(-194,138){ $\left(d\right) $}  \put(-60,40){ $\psi $}
           \caption{PDF of orientation for gyrotactic swimmers of aspect ratio $\mathcal{AR}=9$, swimming speed $V_{s}^*$=1, with respect to: $\left(a\right) $, $\left(b\right) $ and $\left(c\right)$ the three eigenvectors of the deformation tensor; $\left(d\right) $ the local vorticity vector (inset) and the
           direction of gravity. }\label{fig:PDFgyro1}
\end{center}
\end{figure}

We show in figure~\ref{fig:rdfgyro}$(b)$ and $(c)$, and more clearly in the close-up in figure~\ref{fig:rdfgyro}$(d)$, that the maximal clustering of prolate swimmers, yet significantly weaker than that observed for spherical microorganisms,  is for $B\omega_{rms} \approx 2$ whereas it is about 1 for spherical swimmers. {The curve of the scaling exponent versus $B\omega_{rms}$ is more flat in the case of prolate swimmers, whereas a distinct peak is evident for swimmers with $\mathcal{AR}=1$.}
Low values of $B\omega_{rms}$ indicate short re-orientation times and strong torques. In this case, the swimmers tend to align to the vertical direction and swim upwards: an initial uniform distribution will tend to remain as such.  For weak gyrotaxis, large values of $B\omega_{rms}$, particles tend to behave as shown above for non-gyrotactic swimmers and the accumulation is not relevant. Small-scale clustering is therefore occurring for intermediate values of $B\omega_{rms}$ and these optimal values shifts towards longer time scales for prolate swimmers. 
{Comparing figures~\ref{fig:rdfgyro}(a) and \ref{fig:rdf}(a), we note that elongated cells exhibit the weakest clustering, however, they alone have some accumulation when gyrotaxis is turned off.} 
To gain further insight, we examined the behavior of gyrotactic swimmers in laminar uniform shear flows by solving numerically Eq. (\ref{eq4}) for different values of the aspect ratio $\mathcal{AR}$. As discussed below,
 the appearance of Jeffrey-like orbits at larger values of $B\omega_{rms}$, i.e.\ weak reorientations, in the case of prolate swimmers explains the shift in the values of $B\omega_{rms}$ where the largest clustering is observed.

\begin{figure}
\begin{center}
                \includegraphics[width=0.48\textwidth]{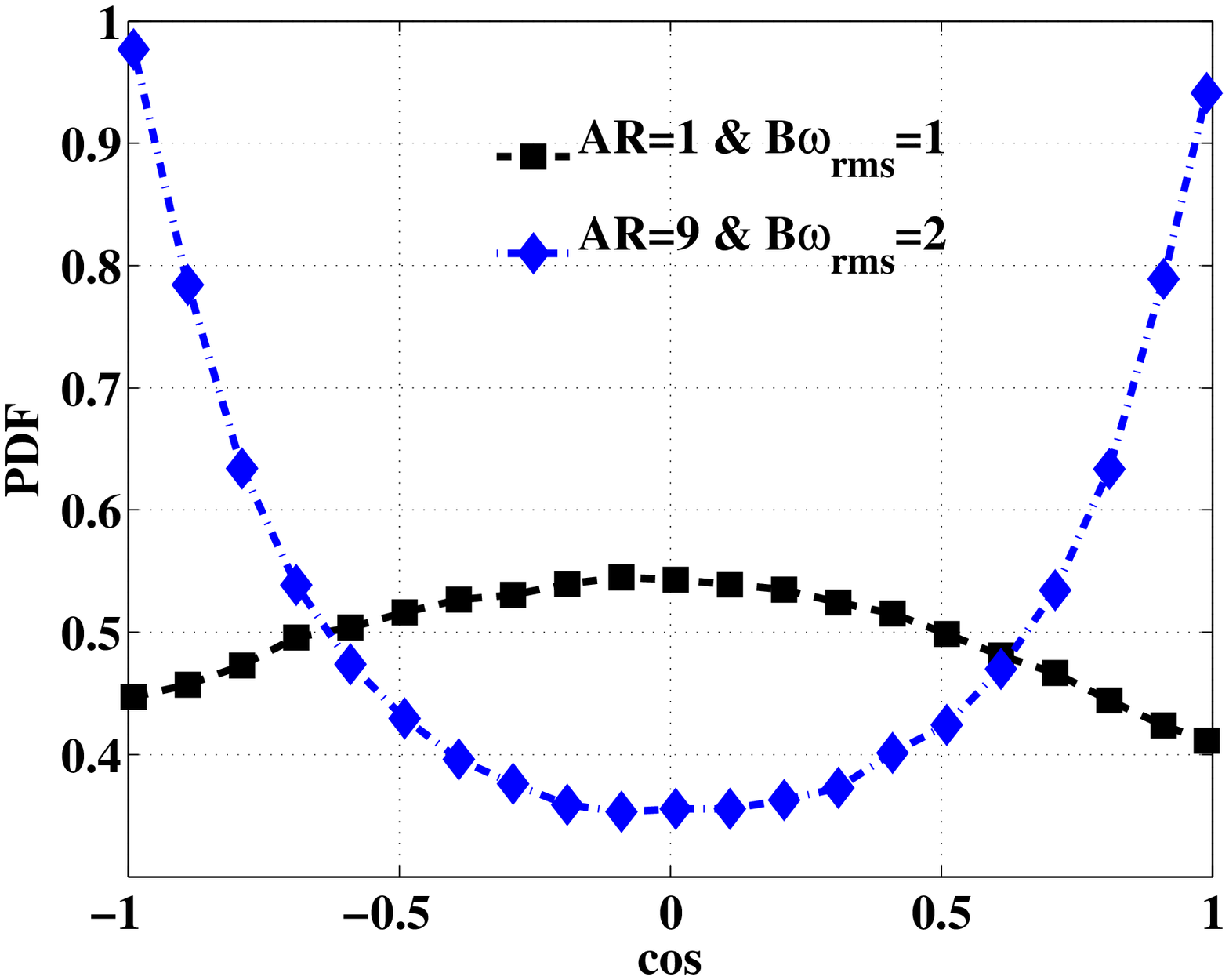}  \put(-192,138){ $\left(a\right) $} \put(-79,2){$\theta_1$}
              \hspace{10pt}  \includegraphics[width=0.48\textwidth]{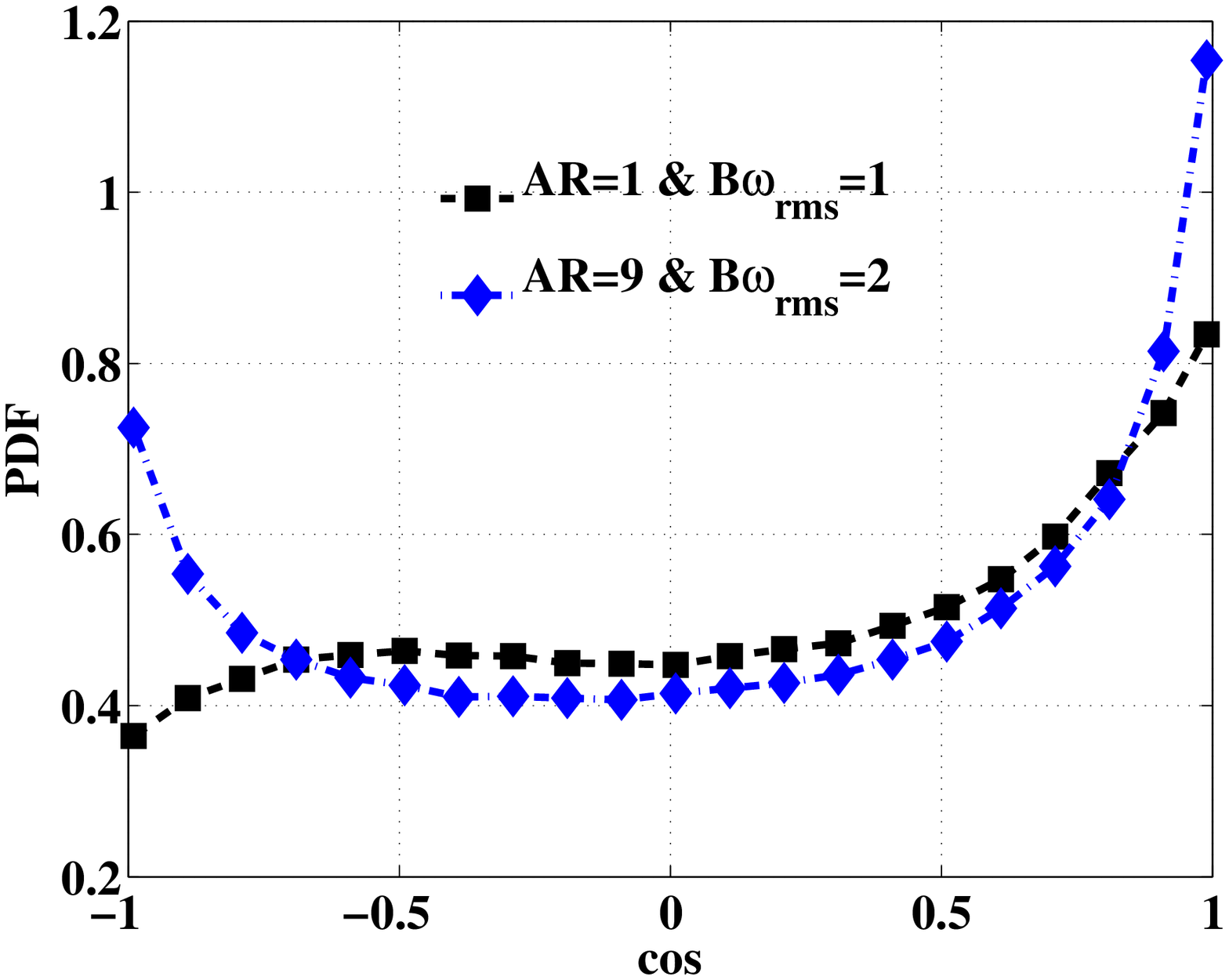}  \put(-194,138){ $\left(b\right) $} \put(-79,2){$\theta_2$}

                \includegraphics[width=0.48\textwidth]{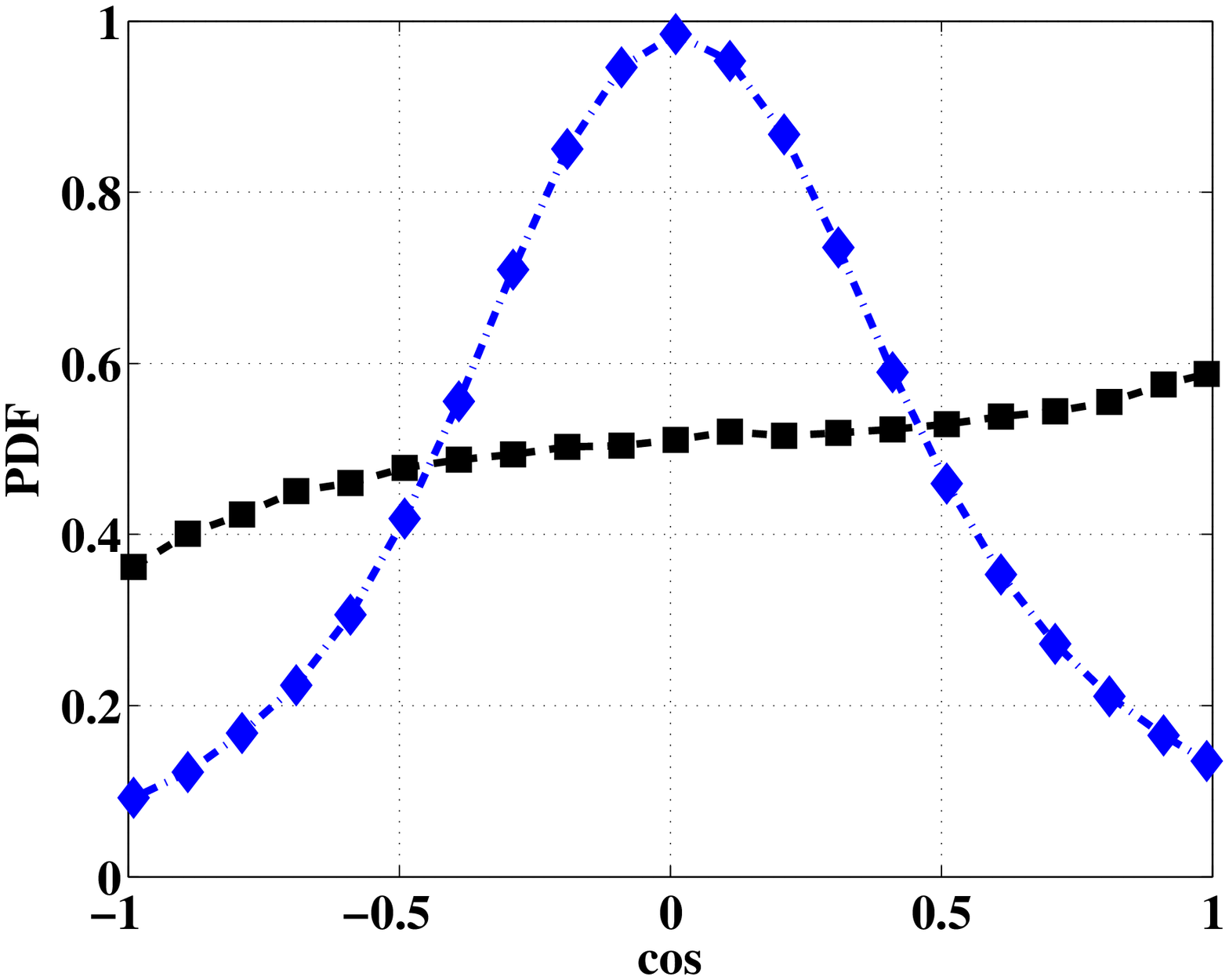}  \put(-192,138){ $\left(c\right) $} \put(-79,2){$\theta_3$}
               \hspace{10pt}  \includegraphics[width=0.48\textwidth]{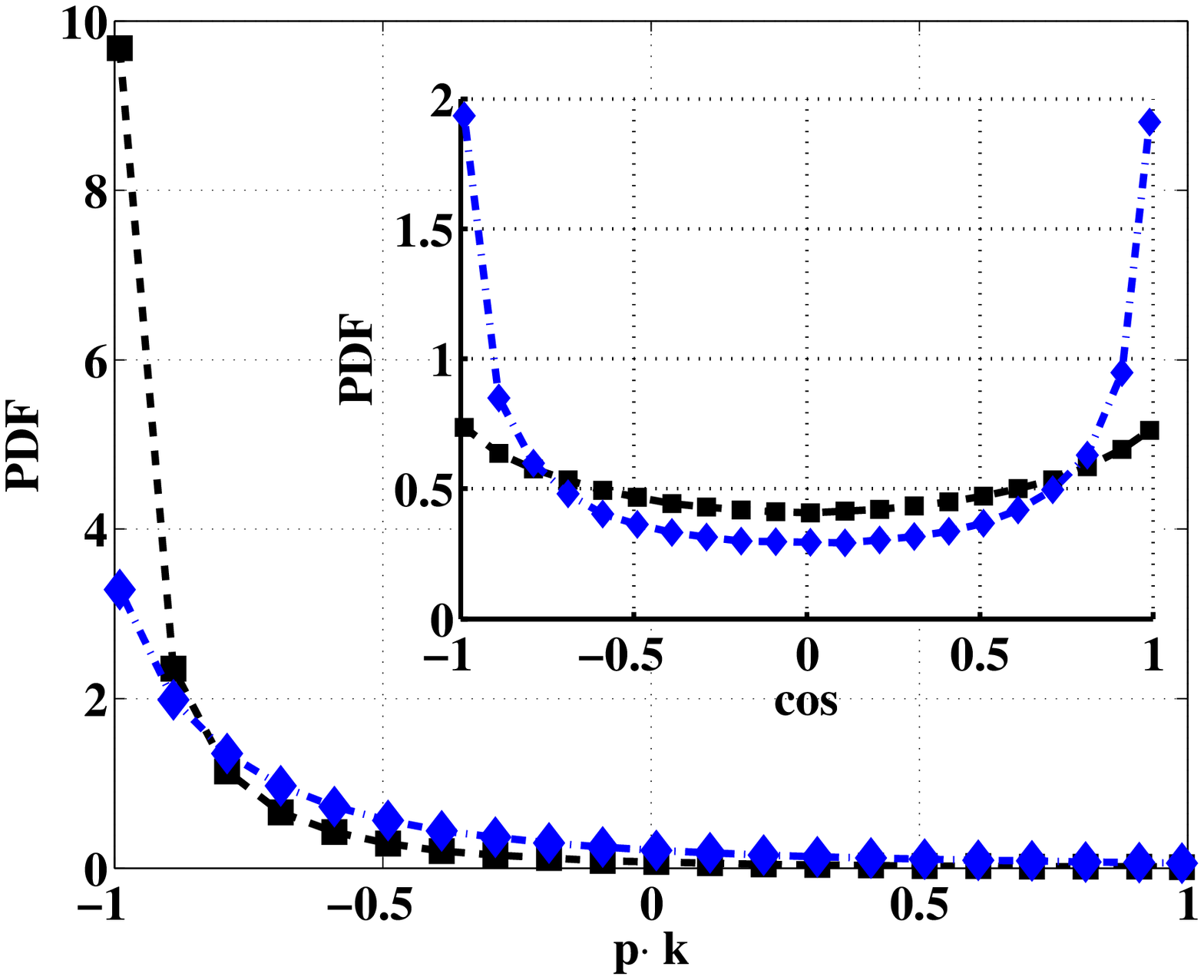}  \put(-194,138){ $\left(d\right) $}  \put(-60,40){ $\psi $}
                         \caption{PDF of orientation for the most accumulating gyrotactic swimmers, $\mathcal{AR}=1, 9$, swimming speed $V_{s}^*$=1, with respect to: $\left(a\right) $, $\left(b\right) $ and $\left(c\right)$ the three eigenvectors of the deformation tensor; $\left(d\right) $ the local vorticity vector (inset) and the
           direction of gravity. }\label{fig:PDFgyro2}
\end{center}
\end{figure}

To further document the decrease of the clustering for prolate swimmers we display in 
figures~\ref{fig:PDFgyro1} and~\ref{fig:PDFgyro2}  the mean orientation of the gyrotactic micro-organisms with respect to the underlying flow field. We keep $\mathcal{AR}=9$ and swimming speed the same and investigate the influence of $B\omega_{rms}$ in figure~\ref{fig:PDFgyro1}. 
The results for weak gyrotaxis, large values of $B\omega_{rms}$, resemble those in figure~\ref{fig:PDF_nogyro}, whereas the orientation with respect to the strain and vorticity fields tends to be more uniform for strong gyrotaxis, where the cell behavior is dominated by gravity (see~\ref{fig:PDFgyro1}d).
In all cases,  peaks of the PDFs occur when the orientation is at 0 and 180 degrees with respect to the direction associated to $\lambda_1$.
The distribution of the angles between the swimming direction and the second direction of the strain tensor becomes asymmetric when increasing the strength of the swimming bias. As for non-gyrotactic swimmers, the orientation of particles is more likely to be normal to the third eigendirection of the strain; this tendency increases as $B\omega_{rms}$ increases.  
In the inset of figure~\ref{fig:PDFgyro1}d) we see that the alignment with the vorticity vector also increases as the stability of the cell relative to that of the flow increases.
Finally we consider the relative orientation with respect to the direction of gravity. As expected, particles tend to align more in the vertical direction as $B\omega_{rms}$ decreases. 
As mentioned above, for strong gyrotaxis swimmers tend to move upwards keeping a uniform distribution, while for lower magnitudes of the gyrotactic torque they are mainly oriented by the flow vorticity and strain.

In figure~\ref{fig:PDFgyro2}, we investigate the alignments of gyrotactic swimmers for the values of $B\omega_{rms}$ giving the maximal clustering,  $B\omega_{rms}=2$ for $\mathcal{AR}=9$, and $B\omega_{rms}=1$ for $\mathcal{AR}=1$.
Prolate swimmers align with the  direction of strain defined by $\lambda_1$ 
whereas  spherical particles are more likely to be normal to it. Conversely, prolate are orthogonal to the third eigendirection while spherical swimmers tend to be more uniformly distributed.  Further, prolate swimmers tend to align to the vorticity vector more than their spherical counterpart.
The shape of the swimmers also affect the orientation with respect to the gravity vector with the spherical swimmers more likely to swim parallel to gravity at given orientation time. Recalling that the spherical swimmers have larger aggregation than those elongated,  we find the orientation of spherical particles to be more strongly dominated by  gravity and less by the local shear. 

To quantify the possible ecological implications of clustering induced by biased swimming in turbulence, we estimate the collision rate of micro swimmers.
To this end, we follow the approach adopted in the study of inertial particles. The particle collision rate  is proportional to clustering (RDF) at a distance corresponding approximatively to one particle diameter and to the mean particle concentration weighted by the probability of particles having negative (approaching) relative velocity \cite[see][]{sundaram97,gualtieri2012statistics,sardina2012wall}: 
\begin{equation}
N_{c}=\pi c^2 \sigma ^2  g_{0} \left(\sigma \right) \langle \delta  v_{p}\left(\sigma \right) |\delta v_{p} <0 \rangle. 
\label{eq_colli}
\end{equation}
In the expression above $N_{c}$ is the total collision rate, c the mean concentration, $g_{0}$ the RDF,  $\sigma$  the length scale characteristic of the collision and $\langle \delta v_{p} \left(\sigma \right) |\delta v_{p} <0 \rangle$ the spherical average of the mean relative velocity of two microorganisms. A similar relation is adopted in \cite{kjorboe99,kiorboe2008}, where a uniform organism distribution is assumed. This observable, used for  inertial particles in turbulence where accumulation occurs, enables us to take into account the combined effect of clustering and swimming.

\begin{figure}
\begin{center}
                \includegraphics[width=0.48\textwidth]{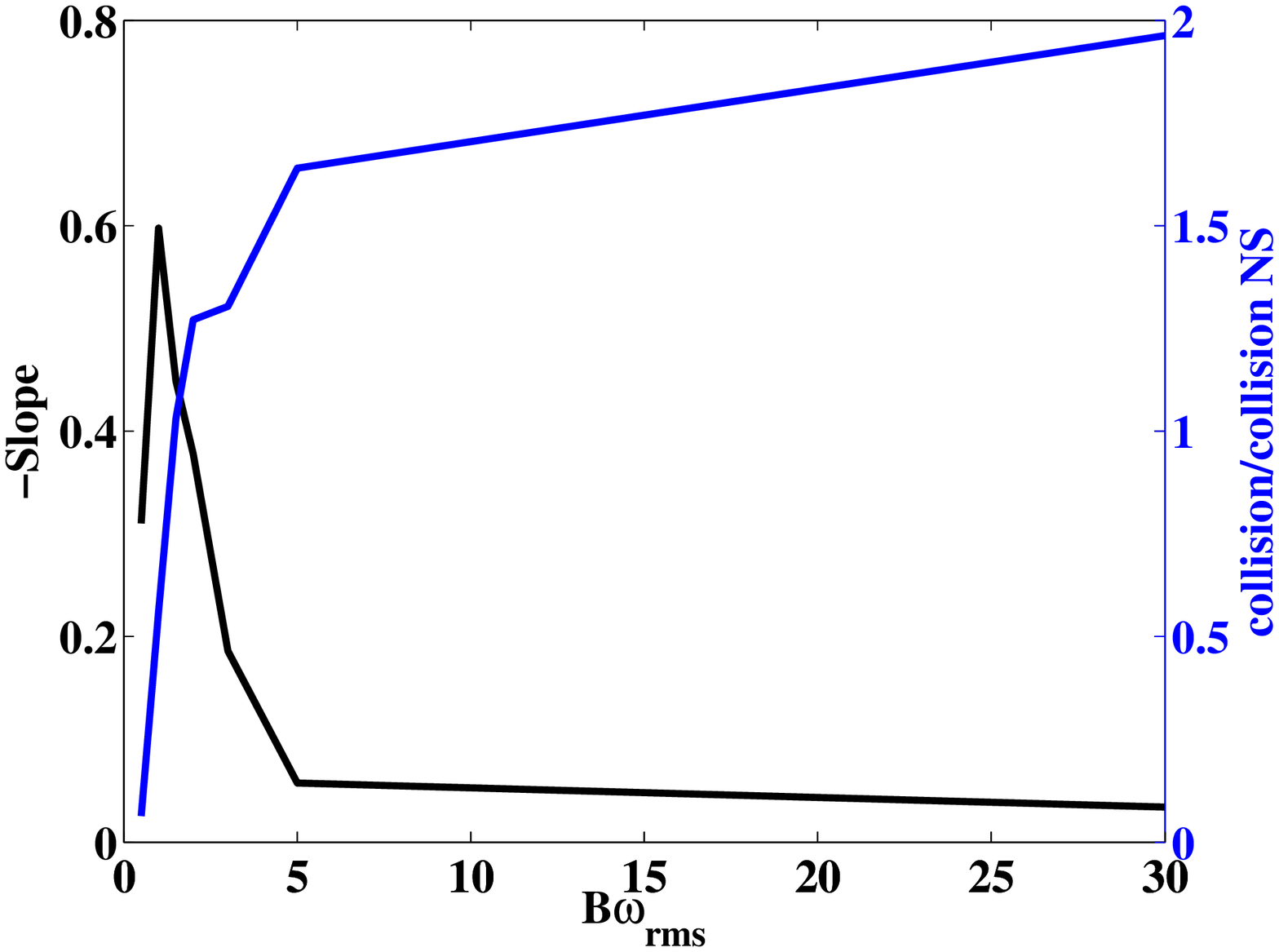}  \put(-193,133){ $\left(a\right) $} 
              \hspace{10pt}  \includegraphics[width=0.48\textwidth]{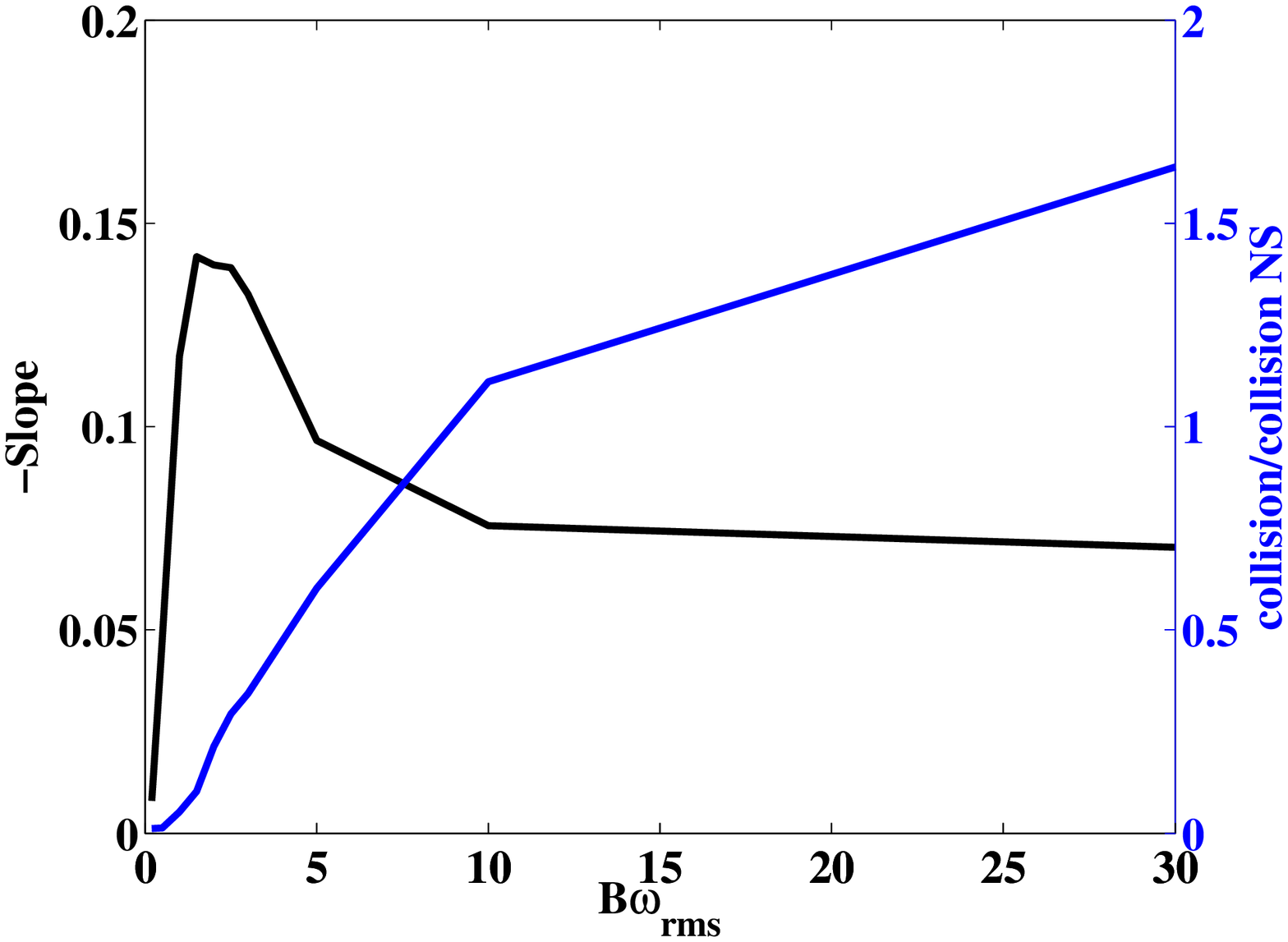}  \put(-193,133){ $\left(b\right) $} 

               \caption{Collision rate and slope of the RDF at small separations versus the non-dimensional re-orientation time $B\omega_{rms}$ for swimming speed $V_{s}^*$=1. $\left(a\right)$ Aspect ratio $\mathcal{AR}=1$ and $\left(b\right) $ Aspect ratio $\mathcal{AR}=9$.}\label{fig:colli}
\end{center}
\end{figure}

The collision rates computed for the different values of the gyrotactic torque and two values of the cell aspect ratio are reported in figure \ref{fig:colli}. In the plot, the collision rates are divided by the value pertaining the case of non-swimming organisms without gyrotaxis (denoted as collision NS); the value $\sigma=0.01\eta$ is used to define a distance of approach. The data for these low values of separation have been extrapolated by assuming a power law for the collision rate:
$N_c \propto g_0 (\sigma) \sigma^2  \langle \delta  v_{p}\left(\sigma \right) |\delta v_{p} <0 \rangle$ is a power law at small scales since $\sigma^2$ is a power law by definition and $ \langle \delta  v_{p}\left(\sigma \right) |\delta v_{p} <0 \rangle$ is known to be well approximated by a power law (classical Kolmogorov theory) in the viscous range. Results obtained with values to $\sigma=0.5\eta$ and thus without extrapolation do not show significant  qualitative differences.
The slope of the RDF is also reported in the figure to display the peak in clustering for the same configurations.
The data show that 
the collision rates increases with $B\omega_{rms}$ reaching {values about twice as large and 50\% larger than those for passive tracers
It can be seen in the figure that the magnitude of the gyrotactic torque is the most important factor affecting the collision rate, more than the clustering. }Surprisingly, the most clustered microorganisms show a reduced level of collisions.  This can be explained by the fact that for low values of $B\omega_{rms}$ the microorganisms, although close to each other, tend to swim in the same direction and thus the collision rate decreases due to the low values attained by the factor $\langle \delta  v_{p}\left(\sigma \right) |\delta v_{p} <0 \rangle$. 
Our results therefore suggest that other (sensorial) mechanisms should be active at the micro scale, sub-Kolmogorov scale, to enhance encounters between those microorganisms that turbulent motions may have brought close to each other.

\subsection{Equilibrium orientation in laminar flows}

 To try to explain the results presented above for a fully three-dimensional time-dependent turbulent flow we examine the solution of the equation governing the swimmer orientation, eq. (\ref{eq4}), for different values of the aspect ratio $\mathcal{AR}$. An analytical steady-state solution can be found for the case of spherical swimmers and is discussed in \cite{Pedley87,thorn}. The cells have a stable orientation parallel to gravity for low values of $B \Omega$, where $\Omega$ is the vorticity of the background laminar flow. This regime equilibrium rotate towards the direction of the vorticity vector for large values of $B \Omega$. A stable solution exists for all values of the non-dimensional parameter $B \Omega$ if the vorticity vector is not orthogonal to the the gravitational force, or equivalently, the gravitational torque is not parallel to the vorticity vector. In this case, an equilibrium solution exists only if $B \Omega<1$; above this value the cells tumble (the limit for infinite B, no gyrotaxis, being the Jeffrey orbit).

\begin{figure}
\begin{center}
      \includegraphics[width=0.44\textwidth]{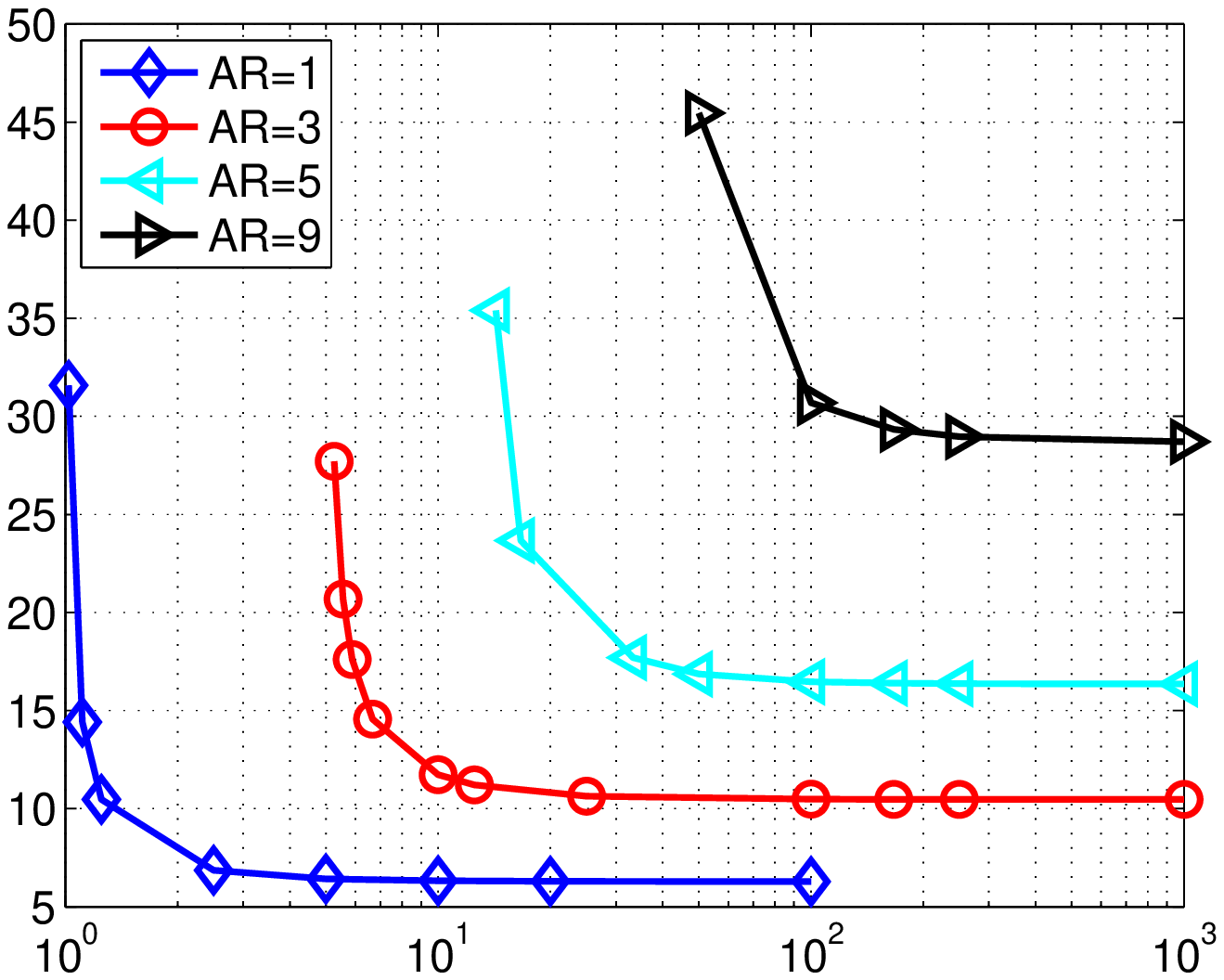}\hspace{1cm}
      \includegraphics[width=0.44\textwidth]{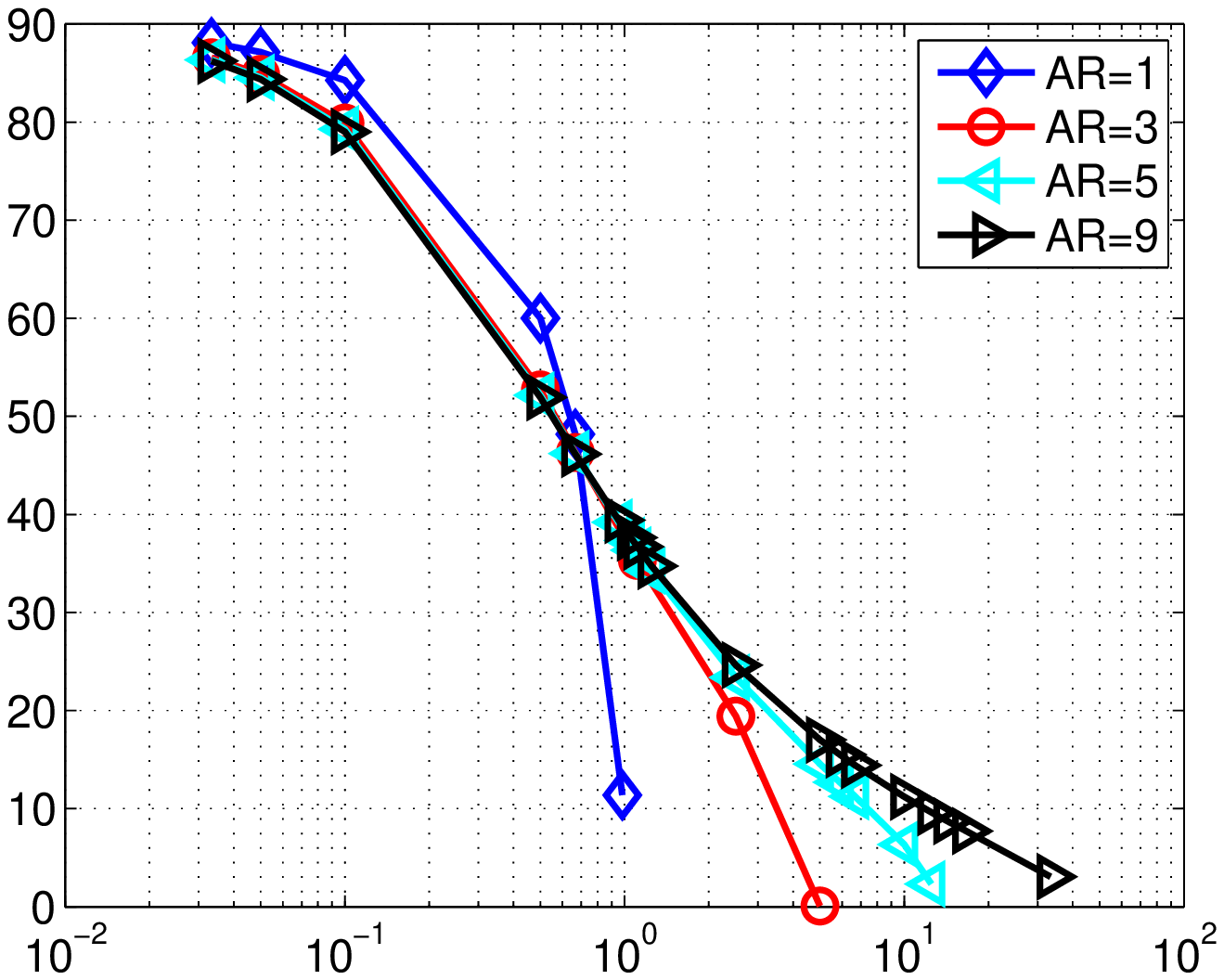}
      \put(-390,125){ $\left(a\right) $}
      \put(-390,60){\large $T\Omega$}
      \put(-296,-11){\large $B \Omega$}
      \put(-90,-11){\large $B \Omega$}
      \put(-180,60){\large $\Theta $}
       \put(-190,125){ $\left(b\right) $}
    \caption{Orientation and tumbling of gyrotactic cells in homogenous shear with gravity in the direction of the velocity gradient. (a) Period of tumbling versus $B \Omega$, with $\Omega$ the shear flow vorticity. (b) Equilibrium angle $\Theta$ in the $x-z$ plane, with $x$ the flow direction and $z$ the direction of gravity.}\label{fig:a90}
\end{center}
\end{figure}

Here, we compute numerical solutions of the time-dependent eq. (\ref{eq4}) and determine the effect of the swimmer shape on the equilibrium orientation, as well as on the appearance of tumbling. We first analyze the case of homogenous shear flow in the plane parallel to gravity, with gravity in the direction of the velocity gradient ($u_x= \Omega z$). Figure~\ref{fig:a90}$(a)$ reports the period of tumbling versus $B \Omega$ for the 4 different values of the aspect ratio considered in the turbulent cases above, whereas~\ref{fig:a90}$(b)$ displays the angle at equilibrium in the $x-z$ plane, with $x$ the flow direction and $z$ the direction of gravity. Note that the two plots are complementary as either tumbling or an equilibrium angle is observed for each value of $B \Omega$ considered.
As swimmers are more and more elongated, the tumbling motion is observed to start at larger values of $B \Omega$, that is for long reorientation times. 
The period of tumbling decreases and approaches that of the Jeffrey orbit as $B \Omega \to \infty$. Below the threshold value for the occurrence of tumbling, the equilibrium angle increases from zero, swimmers almost aligned with the flow, to $90^o$ as $B \Omega \to 0$ , e.g.\ swimmers parallel to gravity for strong gyrotaxis.
The figure clearly shows that the transition between gravity-dominated swimming and tumbling-motions shifts to larger $B \Omega$ for elongated swimmers.

\begin{figure}
\begin{center}
      \includegraphics[width=0.44\textwidth]{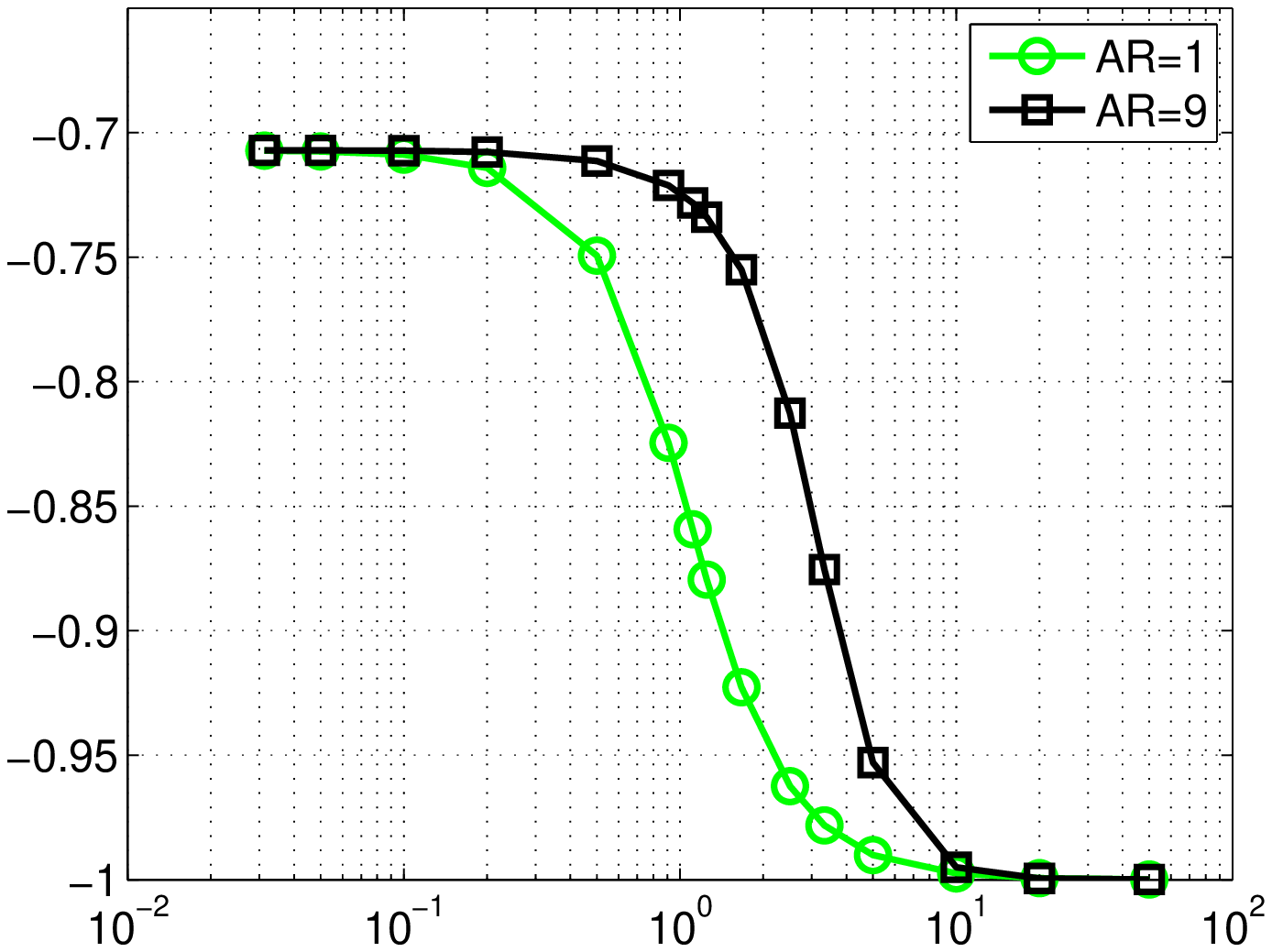}\hspace{1cm}
      \includegraphics[width=0.44\textwidth]{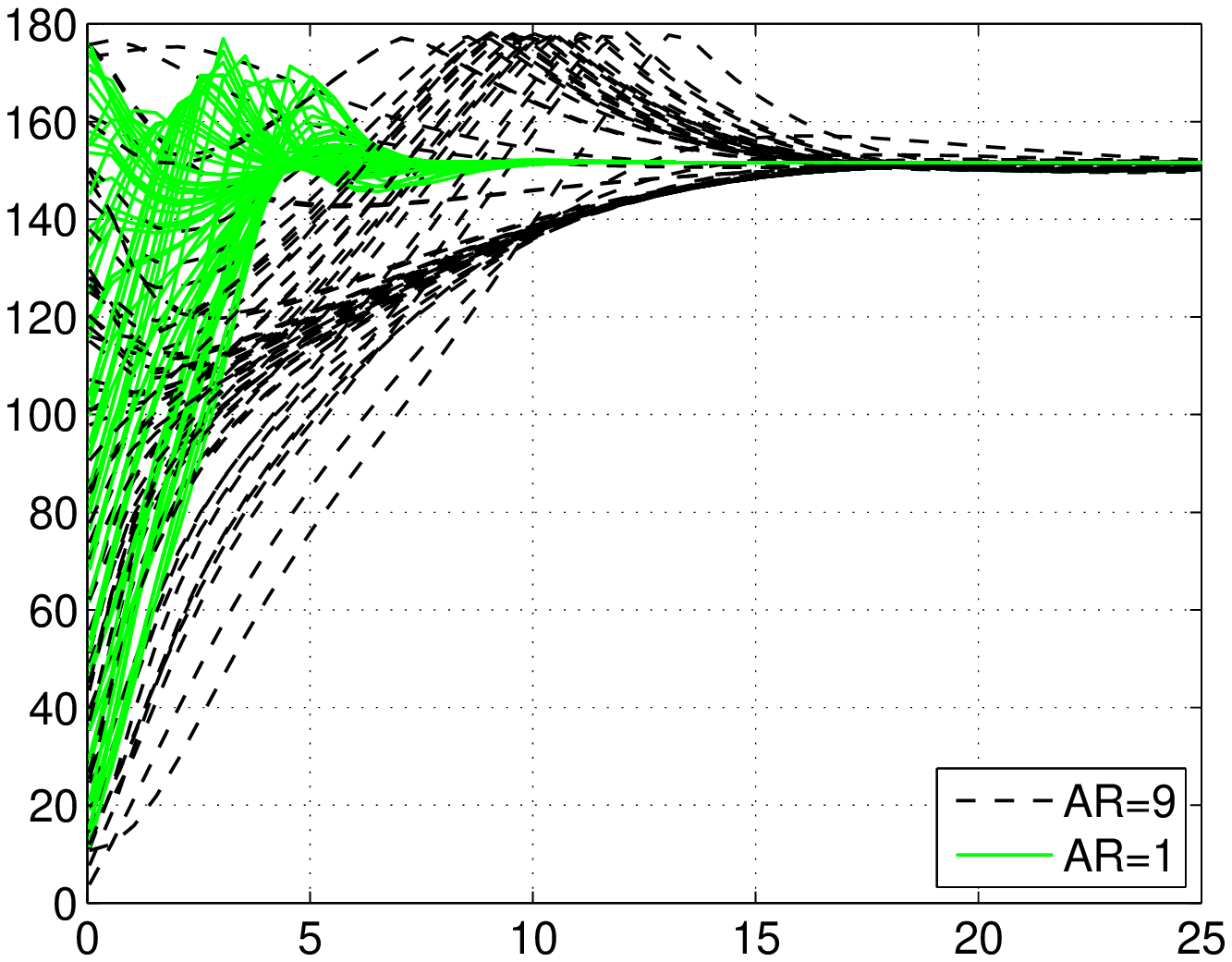}
      \put(-385,120){$\left(a\right) $}
      \put(-390,60){\large $p \cdot \Omega $}
      \put(-292,-14){\large $B \Omega$}
      \put(-85,-14){\large $t \Omega$}
      \put(-180,60){\large $\Theta $}
       \put(-190,120){ $\left(b\right) $}
    \caption{Orientation of gyrotactic cells in homogenous shear with gravity inclined by $45^o$ with respect to the vorticity vector. (a) Equilibrium angle expressed as the scalar product between the regime solution and the vorticity vector $p \cdot \Omega $. (b) 
    Time evolution of the cell orientation for different initial orientations and values of $B \Omega $ during the transition between gravity-dominated swimming and vorticity-dominated swimming (same final orientation).}\label{fig:a45}
\end{center}
\end{figure}

Next we consider the case of vorticity and gravity vector forming an angle of $45^o$, as in \cite{thorn}. In this case, no tumbling is observed and the cells tend to orient parallel to the gravitational force for small values of $B \Omega $, whereas they are aligned with the vorticity vector for large $B \Omega $. This is displayed in figure~\ref{fig:a45}$(a)$ where we report the scalar product between the cell orientation at the final steady state and the flow vorticity, with $\pm \sqrt{2}/2$ indicating the direction of gravity. Again, we note that the transition between gravity-dominated swimming and flow-dominated orientation, the regime where the largest clustering is observed in turbulence, is shifted to large values of $B \Omega $ for elongated swimmers. While this accounts for the shift of the maximum accumulation to larger values of $B \omega_{rms}$ in the turbulent case documented above, it does not help to explain the reduced clustering. To this aim, we report in figure~\ref{fig:a45}$(b)$ { the time evolution of the cell orientation for different initial orientations 
and the values 
$(\mathcal{AR}, B \Omega)= (1 , 0.625)$ and $(\mathcal{AR}, B \Omega)= (9 , 1.66)$ right during the transition between gravity-dominated swimming and vorticity-dominated swimming (same final orientation). Elongated swimmers, $\mathcal{AR}=9$, react more slowly and approach the equilibrium position at a later time, $t \approx 20$, whereas spherical swimmers reach their final equilibrium at $t \approx 10$, where these values correspond approximatively to the mean values computed by varying the initial orientation. The trend shown in the figure (longer transients for elongated cells) is observed in the range $0.25<B\Omega<1.7$ for the spherical swimmers and $0.63<B\Omega<2.5$ for those with $\mathcal{AR}=9$. This range of values is consistent with the values of $B\omega_{rms}$ where maximum clustering is observed in turbulence.} This would imply that in a time-dependent flow elongated swimmers are more unlikely to reach their equilibrium position and their orientation would be determined to a larger extent by the initial condition. As a consequence they would not be able to accumulate like spherical swimmers.

\section{Summary and Discussion}\label{sec:discussion}

In this paper we consider direct numerical simulations of the dynamics of non-spherical swimmers in realistic turbulent flows. 
We demonstrate how the clustering and trapping of swimmers without a preferential alignment observed in simple cellular or vortical flows \cite[e.g.][]{Torney07,Khurana11} 
is significantly reduced in a three-dimensional time-dependent flow and exceeds that of a Poisson distribution for elongated cells, whereas spherical cells remain uniformly distributed.

In addition, we expand on recent studies of small-scale patchiness in the distribution of gyrotactic micro-organisms in vortical or turbulent flows. We confirm that bottom-heavy cells tend to accumulate in downwelling flows, a fact exploited by settling larvae changing the offset of the centers of buoyancy and of gravity to preferentially accumulate in updrafts, favorable for dispersal, or downdrafts, favorable for settlement \citep{grunbaum03}.
In particular, we investigate how the gyrotactic clustering phenomenon in turbulence is modified by the elongation of the ellipsoidal swimmers. The parameters used in this study are in a realistic range and can also be replicated in the laboratory: for typical marine micro-organisms $B=1-6s$, $u_s =100-200 \mu m/s$, thus $u_s / u_\eta \in [0.02:0.4]$, $B \omega_{rms} \in [0.1, 50]$ \citep{Jumars09,Lillo12}.  The shapes of the marine organisms however vary and many phytoplankton species are not spherical, thus the clustering of such organisms in turbulent flows is affected by their individual geometry.

All other conditions equal, we find that clustering is highest for spherical gyrotactic swimmers and decreases the more elongated the swimmers. The orientation of these latter micro-organisms undergoes longer transient phases, and appear to be  more often a consequence of the local shear rate rather than of the local gravity field in a fluctuating flow;
this explains the decreased accumulation. 
Our finding confirms that microorganisms that can actively change their shape, such as \textit{Ceratocorys horrida} \citep[]{Zirbel02}, have an active control mechanism to alter their distribution and favor encounters or uptake.

\section*{Acknowledgements}

Computer time provided by SNIC (Swedish National Infrastructure for Computing) is gratefully acknowledged.  The present work is supported by the Swedish Research Council (VR), by the Linn\'{e} Flow Centre, KTH, and CSC (Chinese Scholarship Council).

\bibliographystyle{jfm}


\end{document}